\documentclass{article}

\usepackage[utf8]{inputenc}
\usepackage[english]{babel}
\usepackage{graphicx}
\usepackage{xcolor}
\usepackage{multirow}
\usepackage{slashbox}
\usepackage{mathrsfs}
\usepackage{amsmath}
\usepackage{amssymb}
\usepackage{float}
\usepackage{tikz}
\usetikzlibrary{decorations.pathreplacing,shapes,arrows,positioning}
\usepackage{setspace}
\usepackage{enumitem}
\usepackage{natbib}
\usepackage{setspace}
\usepackage{lastpage}
\usepackage{authblk}

\begin{document}

\title{An Advanced Computational Scheme for the Optimization of 2D Radial Reflectors in Pressurized Water Reactors}
\author[2]{T. Clerc}
\author[1]{A. ~H\'{e}bert}
\author[2]{H. ~Leroyer}
\author[2]{J.P. ~Argaud}
\author[2]{B. ~Bouriquet}
\author[2]{A. ~Pon\c{c}ot}

\affil[1]{Institut de G\'{e}nie Nucl\'{e}aire,
P.O. Box 6079, Station ``Centre-Ville'', Montr\'{e}al, Qc., Canada, H3C 3A7}
\affil[2]{\'{E}lectricit\'{e} de France, R\&D, SINETICS,
1 Av. du G\'{e}n\'{e}ral de Gaulle, 92141, Clamart, France}

\renewcommand\Authands{ and }

\maketitle
\pageref{LastPage} Pages
\listoftables
\listoffigures
\newpage

\doublespace

\noindent\unskip\textbf{Abstract}
This paper presents a computational scheme for the determination of equivalent 2D multi-group heterogeneous reflectors in a Pressurized Water Reactor (PWR). The proposed strategy is to define a full-core calculation consistent with a reference lattice code calculation such as the Method Of Characteristics (MOC) as implemented in APOLLO2 lattice code. The computational scheme presented here relies on the data assimilation module known as "Assimilation de donn\'{e}es et Aide \`{a} l'Optimisation (ADAO)" of the SALOME platform developed at \'{E}lectricit\'{e} De France (EDF), coupled with the full-core code COCAGNE and with the lattice code APOLLO2. A first validation of the computational scheme is made using the OPTEX reflector model developed at \'{E}cole Polytechnique de Montr\'{e}al (EPM). As a result, we obtain 2D multi-group, spatially heterogeneous 2D reflectors, using both diffusion or $\text{SP}_{\text{N}}$ operators. We observe important improvements of the power discrepancies distribution over the core when using reflectors computed with the proposed computational scheme, and the $\text{SP}_{\text{N}}$ operator enables additional improvements.\\

\clearpage

\section{Introduction}

Pressurized water reactors (PWR) are the mostly used civil nuclear reactor technology in the world. PWRs are called thermal reactor because the reacton of fission is caused by the thermal neutrons. The core, composed of uranium fuel and light water, respectively used as moderator and coolant, releases the power produced by fission. Around the core is placed the reflector, composed mostly of water and stainless steel layers, which slows down the neutrons, reflect them inside the core and protects the core vessel, slows down the neutrons and reflect them inside the core. The physical structure of the reflector is very heterogeneous, and a inaccurate modeling can lead to important azimuthal asymetries on the neutron flux. This is why the modeling of the reflector represents an issue for high-fidelity core calculations using simplified models such as the diffusion equation. At \'{E}lectricit\'{e} De France (EDF) R\&D, the reflector calculation is currently based on the Lefebvre-Lebigot method \citep{refLefLeb}. This method applies to 2-group diffusion calculations, and imposes the ratios $\frac{\text{J}_{\text{1}}}{\Phi_{\text{1}}}$, $\frac{\text{J}_{\text{2}}}{\Phi_{\text{2}}}$ and $\frac{\Phi_{\text{1}}}{\Phi_{\text{2}}}$ to be consistent with a 1D transport calculation at the interface core/reflector. $\text{J}_{\text{i}}$ and $\Phi_{\text{i}}$ are respectively the current and the flux of the energy group i. The diffusion coefficient is then set to an average value within the fissile core. This method leads to the design of an homogeneous reflector. The purpose of our study is to design a computational scheme able to compute multi-group, heterogeneous reflectors, using diffusion or $\text{SP}_{\text{N}}/\text{S}_{\text{N}}$ solvers. The strategy chosen is to select a full-core simplified calculation consistent with a reference transport calculation based on the MOC, in terms of power distribution over an actual core loading pattern. The reference power distribution is obtained with the lattice code APOLLO2 \citep{refAPOLLO2} developed at Comissariat de l'\'{E}nergie Atomique (CEA). All the full-core simplified calculations are carried out with COCAGNE \citep{refCOCAGNE}\citep{refCOCAGNE2}, the new production code developed at EDF. The study relies on the module known as "Assimilation de Donn\'{e}es et Aide \`{a} l'Optimisation (ADAO)" of the platform SALOME \citep{refSALOME}, also developed at EDF and consistent with COCAGNE. This module is based on the data assimilation theory \citep{refAssDonn}, and which integrates COCAGNE. This module is independent from the physical model and can be used to solve general data assimilation problems. A first validation of the computational scheme is presented in Section 3 with the OPTEX reflector model \citep{refAlain} using the full-core simulation code DONJON \citep{refDONJON}, developed at \'{E}cole Polytechnique de Montr\'{e}al (EPM). In this section, we also expose our modeling choices according to the possibilities offered by the data assimilation theory. In Section 4, the results obtained with our computational scheme for diffusion calculations are presented, and in Section 5, the results for $\text{SP}_{\text{3}}$ calculations are presented. As a result, we observe important improvements of the power discrepancies with respect to the reference distribution, which confirms the interest of our computational scheme. We also note an additional improvement of the results when using a $\text{SP}_{\text{3}}$ operator. All the figures of this paper have been published by Clerc in his master thesis \citep{refMoi}.

\section{General context of the study}

\subsection{The study set up}
\begin{figure}[H]

 \begin{tiny}
 \tikzstyle{R}=[rectangle, draw=black, thick, minimum width=0.43cm , minimum height=0.43cm , node distance=0.43cm, text height = 0.5em , fill=gray!10]
 \tikzstyle{u1800}=[rectangle, draw=black, thick , minimum width=0.43cm , minimum height=0.43cm , node distance=0.43cm , text height = 0.5em , fill=white!100]
 \tikzstyle{u2412}=[rectangle, draw=black, thick , minimum width=0.43cm , minimum height=0.43cm , node distance=0.43cm , text height = 0.5em , fill=green!30]
 \tikzstyle{u2416}=[rectangle, draw=black, thick , minimum width=0.43cm , minimum height=0.43cm , node distance=0.43cm , text height = 0.5em , fill=cyan!20]
 \tikzstyle{u2420}=[rectangle, draw=black, thick , minimum width=0.43cm , minimum height=0.43cm , node distance=0.43cm , text height = 0.5em , fill=cyan!50]
 \tikzstyle{u3100}=[rectangle, draw=black, thick , minimum width=0.43cm , minimum height=0.43cm , node distance=0.43cm , text height = 0.5em , fill=orange!50]
 \tikzstyle{u3108}=[rectangle, draw=black, thick , minimum width=0.43cm , minimum height=0.43cm , node distance=0.43cm , text height = 0.5em , fill=yellow!30]
 \tikzstyle{u3116}=[rectangle, draw=black, thick , minimum width=0.43cm , minimum height=0.43cm , node distance=0.43cm , text height = 0.5em , fill=red!40]
 \tikzstyle{u3124}=[rectangle, draw=black, thick , minimum width=0.43cm , minimum height=0.43cm , node distance=0.43cm , text height = 0.5em , fill=purple!50]

 \begin{center}
  \begin{tikzpicture}
   \node[name=1,u1800] at (6,0) {1};
   \node[name=2,u2416,right of=1] {2};
   \node[name=3,u1800,right of=2] {3};
   \node[name=4,u2416,right of=3] {4};
   \node[name=5,u1800,right of=4] {5};
   \node[name=6,u3124,right of=5] {6};
   \node[name=7,u1800,right of=6] {7};
   \node[name=8,u2420,right of=7] {8};
   \node[name=9,u3100,right of=8] {9};
   \node[name=34,R,right of=9] {R};
   \node[name=35,R,right of=34] {R};

   \node[name=10,u1800,above of=2] {10};
   \node[name=11,u2412,right of=10] {11};
   \node[name=12,u1800,right of=11] {12};
   \node[name=13,u2416,right of=12] {13};
   \node[name=14,u1800,right of=13] {14};
   \node[name=15,u2420,right of=14] {15};
   \node[name=16,u3100,right of=15] {16};
   \node[name=17,u3100,right of=16] {17};
   \node[name=36,R,right of=17] {R};
   \node[name=37,R,right of=36] {R};

   \node[name=18,u1800,above of=11] {18};
   \node[name=19,u2412,right of=18] {19};
   \node[name=20,u1800,right of=19] {20};
   \node[name=21,u2412,right of=20] {21};
   \node[name=22,u1800,right of=21] {22};
   \node[name=23,u3108,right of=22] {23};
   \node[name=38,R,right of=23] {R};
   \node[name=39,R,right of=38] {R};
   \node[name=40,R,right of=39] {R};

   \node[name=24,u1800,above of=19] {24};
   \node[name=25,u2412,right of=24] {25};
   \node[name=26,u1800,right of=25] {26};
   \node[name=27,u3116,right of=26] {27};
   \node[name=28,u3100,right of=27] {28};
   \node[name=41,R,right of=28] {R};
   \node[name=42,R,right of=41] {R};
   \node[name=43,R,right of=42] {R};

   \node[name=29,u1800,above of=25] {29};
   \node[name=30,u2420,right of=29] {30};
   \node[name=31,u3100,right of=30] {31};
   \node[name=44,R,right of=31] {R};
   \node[name=45,R,right of=44] {R};
   \node[name=46,R,right of=45] {R};
   \node[name=47,R,right of=46] {R};

   \node[name=32,u3100,above of=30] {32};
   \node[name=33,u3100,right of=32] {33};
   \node[name=48,R,right of=33] {R};
   \node[name=49,R,right of=48] {R};
   \node[name=50,R,right of=49] {R};
   \node[name=51,R,right of=50] {R};

   \node[name=52,R,above of=33] {R};
   \node[name=53,R,right of=52] {R};
   \node[name=54,R,right of=53] {R};
   \node[name=55,R,right of=54] {R};
   \node[name=56,R,right of=55] {R};

   \node[name=57,R,above of=53] {R};
   \node[name=58,R,right of=57] {R};
   \node[name=59,R,right of=58] {R};
   \node[name=60,R,right of=59] {R};

   \node[name=61,R,above of=58] {R};
   \node[name=62,R,right of=61] {R};
   \node[name=63,R,right of=62] {R};

   \node[name=64,R,above of=62] {R};
   \node[name=65,R,right of=64] {R};

   \node[name=66,R,above of=65] {R};
  \end{tikzpicture}
 \end{center}
 \end{tiny}
\caption{Eighth of core (1 to 33: core; R: reflector)}
\label{fig:figureCoeur1}
\end{figure}
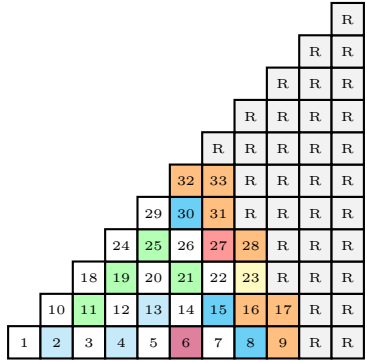

The experimental set up is a eighth of core containing 33 fissile homogeneous assemblies, as described in Figure~\ref{fig:figureCoeur1}. Each assembly is individually processedusing APOLLO2, in fundamental mode approximation. The reflector can be refined or homogenized according to 33 homogeneous assemblies.
In this study, all the core calculation are performed with the core code COCAGNE to compute the power distribution over the core. COCAGNE contains a $\text{SP}_{\text{N}}$ solver based on the Raviart-Thomas finite elements method. The $\text{SP}_{\text{1}}$ solver can be adapted to perform the diffusion calculations presented is this paper. Cubical finite elements, and $8\times8$ elements per assembly are used here. The reference power distribution is recovered from an 281-group and 142,872-region APOLLO2 calculation collapsed to 26 groups and homogenized on each assembly of Figure~\ref{fig:figureCoeur1}, performed with the Method Of Characteristics (MOC). This reference calculation is collapsed and homogenized according to the selected parameters of our computational scheme.
\\[1.0em] 
The strategy of the study is to minimize the discrepancies between a power distribution computed with COCAGNE and the reference APOLLO2 power distribution, by modifying certain parameters of the reflectors, that are called control variables. For diffusion calculations, these control variables are the diffusion coefficients of the reflector, noted $\text{D}_{\text{g,R}_{\text{i}}}$ for the energy group \textit{g} and the reflector region $\text{R}_{\text{i}}$. For $\text{SP}_{\text{N}}$ calculations, the control variables are the P-1 weighted total macroscopic cross-sections. These P-1 weighted total cross sections
are those appearing in the odd-parity $\text{SP}_{\text{N}}$ equations.
\\[1.0em]
As the APOLLO2 calculation doesn't compute diffusion coefficient in the reflector, they will be obtained with the Lefebvre-Lebigot method, or recovered from the 26-group reference calculation. In this case, they will be collapsed, such as \mbox{$\text{D} =  \frac{1}{3\int_{\text{E}} \Sigma_{\text{tr}}\text{(E)dE}}$}, where $\Sigma_{\text{tr}}$ is the macroscopic total cross-section, and homogenized such as 
\mbox{$\text{D} = \int_{\text{V}} \text{D(V)dV}$}, where V is the volume of an assembly. Here, the macroscopic transport cross section is defined as:
\begin{equation}
\Sigma_{tr,g,R_{i}} = \Sigma_{g,R_{i}}-\sum_{h=1}^{G}\Sigma_{s,1,g\leftarrow h,R_{i}}
\end{equation}
The diffusion coefficient calculation strategy will be explained in Section 2.4.

\subsection{The data assimilation theory}
The data assimilation theory was first introduced in the field of meteorological studies, to recover a temperature field or to perform weather forecasts. In the module ADAO, it can be used to solve any data assimilation problem, such as those appearing in the fields of mechanics or neutronics. At EDF R\&D, the module is used for field recovering applied to neutronic studies, or for parameters studies. For instance, it is used to study the influence of a measurement points network in a reactor core on the reconstructed flux \citep{refADAO1}, or to set up an optimal measurement network to recover the best reconstruction of a fission rate distribution \citep{refADAO2}. Here, we use the parameters study functions of ADAO.
The data assimilation theory is used to recover the "true state" $\textbf{x}^{\textbf{t}}$ of a system, by gathering all the information available on this system, in terms of physical models, observations and error statistics. This enables to search the true state of systems poorly observed or with approximated physical models. However, we will only obtain an estimate of the true state, that is to say here the result of the minimization of the functional in Eq~\ref{eq:foncAssDonn}. The first guess of this true state is $\textbf{x}^{\textbf{b}}$, and it is associated to its error, defined by $\textbf{B} = \mathbb{E}[(\textbf{x}^{\textbf{b}}-\textbf{x}^{\textbf{t}})\cdot(\textbf{x}^{\textbf{b}}-\textbf{x}^{\textbf{t}})^{T}]$. The actual measurement on the system are noted $\textbf{y}^{\textbf{0}}$. The observation operator is noted \textit{H}, and the observations are then defined by $H(\textbf{x}^{\textbf{t}})$. When \textit{H} is linearized, it is noted \textbf{H} and the observations are then defined by $\textbf{H}\textbf{x}^{\textbf{t}}$. These observations are associated with their error matrix: $\textbf{R} = \mathbb{E}[(\textbf{y}^{\textbf{0}}-\textbf{H}\textbf{x}^{\textbf{t}})\cdot(\textbf{y}^{\textbf{0}}-\textbf{H}\textbf{x}^{\textbf{t}})^{T}]$. The data assimilation strategy is to minimize the discrepancy between the observations and the actual measurements of a system. The most common approach when the observation operator is linear is called Best Linear Unbiased Estimator (BEST) \citep{refAssDonn}. When the observation is not linear, an equivalent approach called 3D-VAR consists in minimizing a functional based on the least squares. The common least-squares functional is weighted to take into account the error on the observations \textbf{R}, and a contribution of the background is added in order to regularize the variations of the functional. We then obtain the functional \textit{J}:


\begin{equation}
J(\textbf{x}) = \frac{1}{2}(H(\textbf{x})-\textbf{y}^{\textbf{0}})^{T}\textbf{R}^{-1}(H(\textbf{x})-\textbf{y}^{\textbf{0}})
 + \frac{1}{2}(\textbf{x}-\textbf{x}^{\textbf{b}})^{T}\textbf{B}^{-1}(\textbf{x}-\textbf{x}^{\textbf{b}})
\label{eq:foncAssDonn}
\end{equation}

Where \textbf{x} is the state vector, containing the current minimization parameters. The two approaches are equivalent if the problem is linear and the error distribution on the observation is a Gaussian \citep{refAssDonn}, which is the most common hypothesis. In ADAO, all the tools to solve a data assimilation problem are offered, and the physical models and measurements must be imported form previously designed python scripts. 

\subsection{Application to our study}
In this study, the state vector of our system contains the diffusion coefficients of each energy group in each zone of the reflector, in the case of diffusion calculations, such as \textbf{x} = \textbf{D}. In the case of $\text{SP}_{\text{N}}$ calculations, the state vector contains P-1 weighted macroscopic total cross sections of each energy group in each zone of the reflector such as $\textbf{x} = \Sigma_{\rm \textbf{P1}}$. The size of \textbf{x} depends on the modeling choices in terms of energy mesh an spatial mesh in the reflector. The measurements of the system are the values of the reference power in each of the 33 fissile assemblies of Figure~\ref{fig:figureCoeur1} ($\text{P}^{\text{*}}_{\text{i}}, \: j \in \text{[1,33]}$), normalized on the eighth of core, such as: 

\begin{equation}
\textbf{y}^{\textbf{0}} = \left(\frac{P^{*}_{j}}{\sum_{k=1}^{33} P^{*}_{k}}, \: j \in [1,33]\right)
\label{eq:measurements}
\end{equation}

The observations of the system are the values of the power computed by COCAGNE in each assembly of the eighth of reactor ($\text{P}_{\text{j}}(\textbf{x}), \: j \in \text{[1,33]}$), normalized on the eighth of reactor, such as:

\begin{equation}
H(\textbf{x}) = \left(\frac{P_{j}(\textbf{x})}{\sum_{k=1}^{33} P_{k}(\textbf{x})}, \: j \in [1,33]\right)
\label{eq:observations}
\end{equation}

The observation operator H is clearly non-linear, and the 3D-VAR approach is then chosen in this study. Moreover, the observations are supposed to be untied, and the reflector parameters also. The error matrix are then diagonals, such as \textbf{B} = diag($\sigma_{\text{B,i}}, \: i \in \text{[1,N]}$), where N is the number of reflector parameters. We also define \textbf{R}=diag($\sigma_{\text{R,j}}, \: j \in \text{[1,33]}$). The data assimilation functional is then written as:

\begin{equation}
 J(\textbf{x})  =  \sum_{i=1}^{N} \frac{1}{2\sigma_{B,i}} (x_{i}-x^{b}_{i}) +
  \sum_{j=1}^{33} \frac{1}{2\sigma_{R,j}}\left(\frac{P_{j}(\textbf{x})}{\sum_{k=1}^{33}P_{k}(\textbf{x})}-\frac{P^{*}_{j}}{\sum_{k=1}^{33} P^{*}_{k}}\right)^{2}
\label{eq:foncAssDonn2}
\end{equation}

\begin{table*}[t]
 \caption{Comparisons of initial reflectors}
 \centering
 \begin{tabular}{|c|c|c|c||c|c|}
  \hline
  \multicolumn{2}{|c|}{collapsing diffusion} & \multicolumn{2}{c||}{collapsing transport} & \multicolumn{2}{c|}{Lefebvre} \\
  \multicolumn{2}{|c|}{$\text{D} = \int{\frac{1}{3\Sigma_{\text{tr}}}}$} & \multicolumn{2}{c||}{$\text{D} = \frac{1}{3\int{\Sigma_{\text{tr}}}}$} & \multicolumn{2}{c|}{Lebigot} \\
  \hline
  $\text{max}_{\Delta}$ $(\%)$ & $\text{E}_{\Delta_{\text{abs}}}$ $(\%)$ & $\text{max}_{\Delta}$ $(\%)$ & $\text{E}_{\Delta_{\text{abs}}}$ $(\%)$ & $\text{max}_{\Delta}$ $(\%)$ & $\text{E}_{\Delta_{\text{abs}}}$ $(\%)$ \\
  \hline
  18.7  & 8.6 & 36.5 & 17.9 & 2.5 & 0.7 \\
  \hline
 \end{tabular}
 \label{tab:refInit}
\end{table*} 

The minimization of this functional is carried out using the L-BFGS method \citep{refLBFGS1}\citep{refLBFGS2}. Given that COCAGNE doesn't include a exact gradient computation tool, the gradient of the functional will be computed in ADAO using mesh-corner finite differences.

\subsection{The initial reflector}

The first challenge of this study is to choose an initial reflector. The parameters of this reflectors used for the optimization (diffusion coefficients or macroscopic total cross-sections) is the background of the data assimilation theory. In this section, 2-group diffusion calculations are performed. There are three options to initialize the reflector:\
\begin{enumerate}
  \item[$\bullet$] Collapsing transport: the initial diffusion coefficients are obtained from the 26-group APOLLO2 calculation, which has been collapsed. The diffusion coefficient are then obtained with: $\text{D}=\frac{1}{3\int_{E}\Sigma_{\text{tr}}\text{(E)dE}}$\
  \item[$\bullet$] Collapsing diffusion: the initial diffusion coefficients are first obtained from the 26-group APOLLO2 calculation, and then collapsed: $\text{D}=\int_{E}\frac{\text{dE}}{3\Sigma_{\text{tr}}\text{(E)}}$\
  \item[$\bullet$] Diffusion coefficients obtained from the Lefebvre-Lebigot metod. \
\end{enumerate}
In Table~\ref{tab:refInit}, the average and maximum discrepancies between the computed power and the reference power distribution are presented for each reflector initializing option.
The results are much more satisfying when using the Lefebvre-Lebigot reflector. However, this reflector is only available for 2-group diffusion calculations. The second best option is collapsing diffusion. Given these results, an initialization strategy of the reflector is identified:
\begin{enumerate}
  \item[$\bullet$] When 2-group diffusion calculations are performed, the reflector will be initialized with the Lefebvre-Lebigot method.\
  \item[$\bullet$] For all the other calculations, the reflector will be initialized by collapsing the diffusion ($\text{D}=\int_{E}\frac{\text{dE}}{3\Sigma_{\text{tr}}\text{(E)}}$).\
\end{enumerate}

\section{Validation and modeling choices}

\subsection{The OPTEX reflector model}
The OPTEX reflector model is developed at EPM. It uses the modules of OPTEX \citep{refOPTEX}, integrated in the core code DONJON \citep{refDONJON}, also developed at EPM and based on Raviart-Thomas finite elements. The OPTEX reflector model represents an alternative method to our computational scheme, to compute equivalent 2D radial reflectors in a PWR. It is based on the Generalized Perturbation Theory (GPT) to compute the exact gradients of a functional, which is then minimized using the Parametric Linear Complementary Pivoting method as described in \citep{refAlain}. The OPTEX reflector model is used in this study to perform a first validation of the computational scheme presented here, on simple test cases described in the next paragraph of this paper.

\subsection{Validation of the computational scheme}
In this section, our computational scheme will be compared to the OPTEX reflector model for 2-group or 4-group diffusion calculations, on several test cases. We first consider a one-parameter optimization case. A 2-group homogeneous reflector is computed by optimizing only the fast group diffusion coefficient of the reflector (noted $\text{D}_{\text{1},\text{R}}$). Then the computational scheme is validated for two parameters optimizations cases:\
\begin{enumerate}
  \item[$\bullet$] A 2-group homogeneous reflector is computed by optimizing the two diffusion coefficients (noted $\text{D}_{\text{1},\text{R}}$ and $\text{D}_{\text{2},\text{R}}$).\
    \item[$\bullet$] A 4-group homogeneous reflector is computed by optimizing the two fast diffusion coefficients (noted $\text{D}_{\text{1},\text{R}}$ and $\text{D}_{\text{2},\text{R}}$), and then  the first and the third diffusion coefficients (noted $\text{D}_{\text{1},\text{R}}$ and $\text{D}_{\text{3},\text{R}}$).
	\item[$\bullet$] A 2-group two-region reflector is computed by optimizing only the fast group diffusion coefficient in each region of the reflector (noted $\text{D}_{\text{1},\text{R}_{\text{1}}}$ and $\text{D}_{\text{1},\text{R}_{\text{2}}}$). \
\end{enumerate}

We use cubic finite elements and each assembly is divided into $8\times8$ elements. The functional of the OPTEX model is the root mean square of the discrepancies distribution between the power computed by DONJON and the reference power. In order for the functionals to be equal, we set:

\begin{eqnarray}
\sigma_{R,j} &= 0.5,& \: \forall j \in [1,33] \\
\sigma_{B,i} &\gg 1,& \: \forall i \in [1,N] 
\label{eq:Val1}
\end{eqnarray}

That is equivalent to assume no contribution from the background for the data assimilation problem, and to set an equal contribution of all the assemblies in the active core. The functional \textit{J} will then be written:

\begin{equation}
J(\textbf{x}) = \sum_{j=1}^{33} \left(\frac{P_{j}(\textbf{x})}{\sum_{k=1}^{33}P_{k}(\textbf{x})}-\frac{P^{*}_{j}}{\sum_{k=1}^{33} P^{*}_{k}}\right)^{2}
\label{eq:foncAssDonn3}
\end{equation}

\begin{figure}[H]
 \begin{center}
  \includegraphics[height=4.5cm,width=5cm]{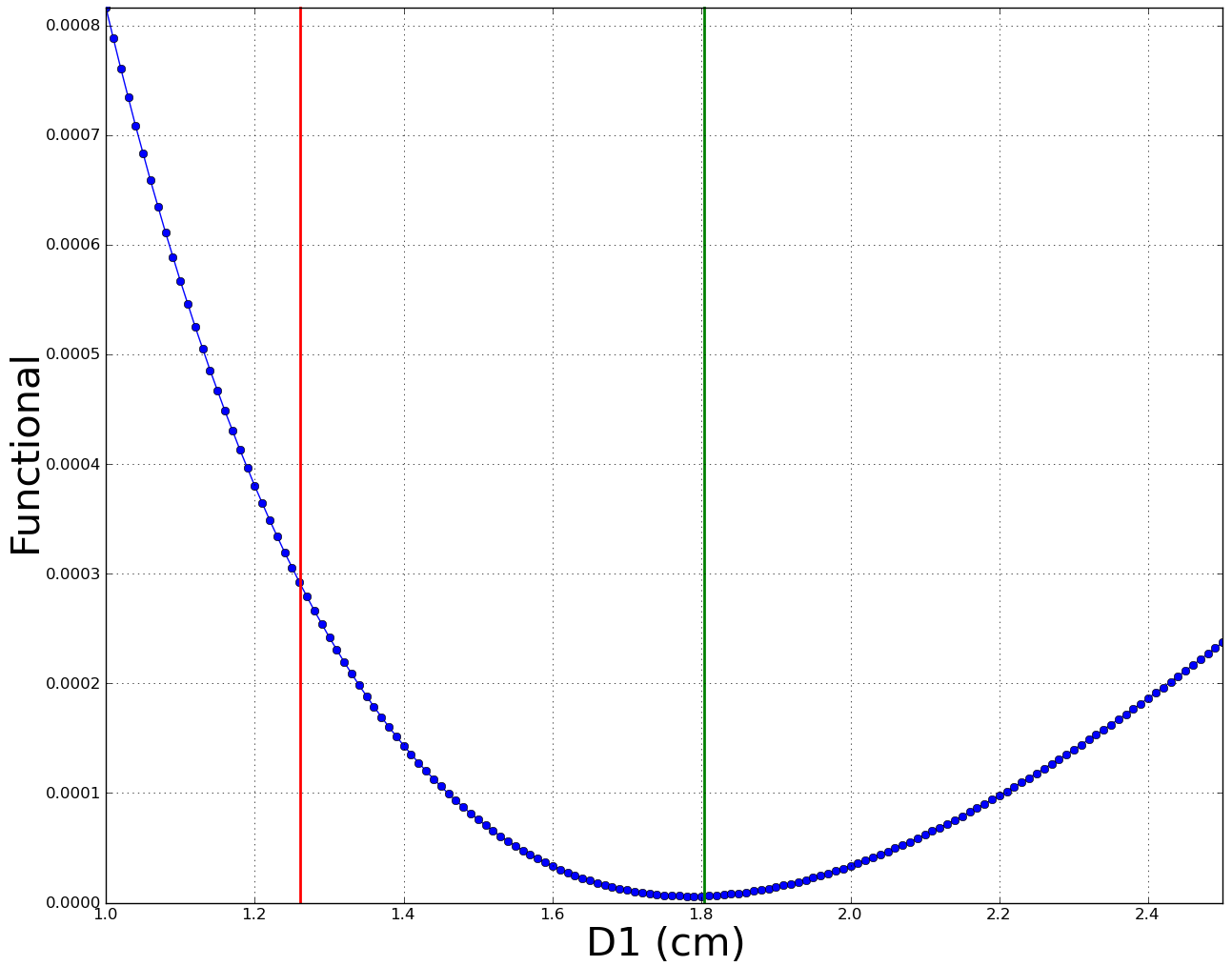}
  \caption{Variations of the functional with $\text{D}_{\text{1},\text{R}}$}
  \label{fig:VarFoncD1}
 \end{center}
\end{figure}

The termination criterion for the two optimization methods is set to $10^{-12}$ on the flux. In fact, the gradients are computed using mesh-corner finite differences and a very strong precision of the flux is needed to compute precise gradients with this method.In the first validation case, the variation of the functional with respect to $\text{D}_{\text{1,R}}$ are presented in Figure~\ref{fig:VarFoncD1}. The positions of the background and the optimum are marked respectively in red and green. The background is higher than the optimum. In fact, the initial reflector leads to very high values of the initial discrepancies, negative at the center of the core and positive at the periphery. When the diffusion coefficient increases, the neutrons leakage toward the reflector increases and the power thus decreases at the periphery of the core. With the normalization effect, the power increases at the center of the core (the values are negative in this region). This causes the decrease of the power discrepancies absolute values over the core.

\begin{figure}[H]
 \begin{center}
  \includegraphics[scale=0.30]{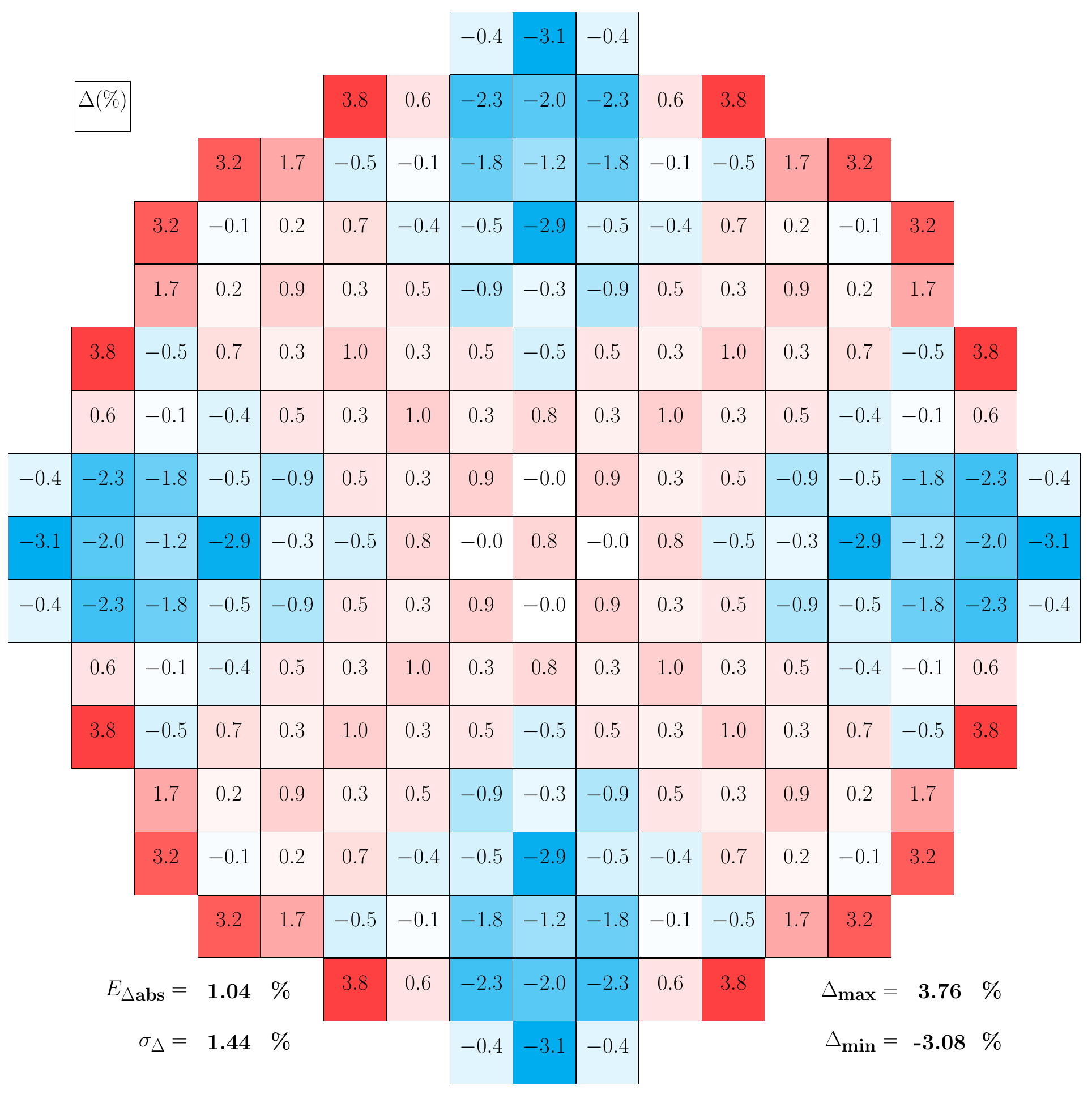}
  \caption{Power discrepancy distribution (validation case 1)}
  \label{fig:puissanceCOCDRAG_D1_ADAO}
 \end{center}
\end{figure}

The discrepancies distribution between the power computed with the core codes and the reference power distribution ($\Delta_{\text{k}}, \: k \in \text{[1,281]}$) is presented in Figure~\ref{fig:puissanceCOCDRAG_D1_ADAO}
. 

This test case is very simple and naturally the discrepancies distributions obtained with the OPTEX reflector model and our computational scheme are the same. The maximum and minimum discrepancies are respectively $\Delta_{\text{max}}$ and $\Delta_{\text{min}}$. $\text{E}_{\Delta_{\text{abs}}}$ is the mean of the distribution and $\sigma_{\Delta}$ is the standard deviation ($\bar{\Delta}$ is the average discrepancy):

\begin{equation}
E_{\Delta_{abs}} = \sum_{k=1}^{281}\frac{|\Delta_{k}|}{281} \: ; \:
\sigma_{\Delta} = \sqrt{\frac{\sum_{k=1}^{281} (\Delta_{i}-\bar{\Delta})^{2}}{281}}
\label{eq:Indicators}
\end{equation}

\begin{figure}[H]
 \begin{center}
  \includegraphics[height=5cm,width=7cm]{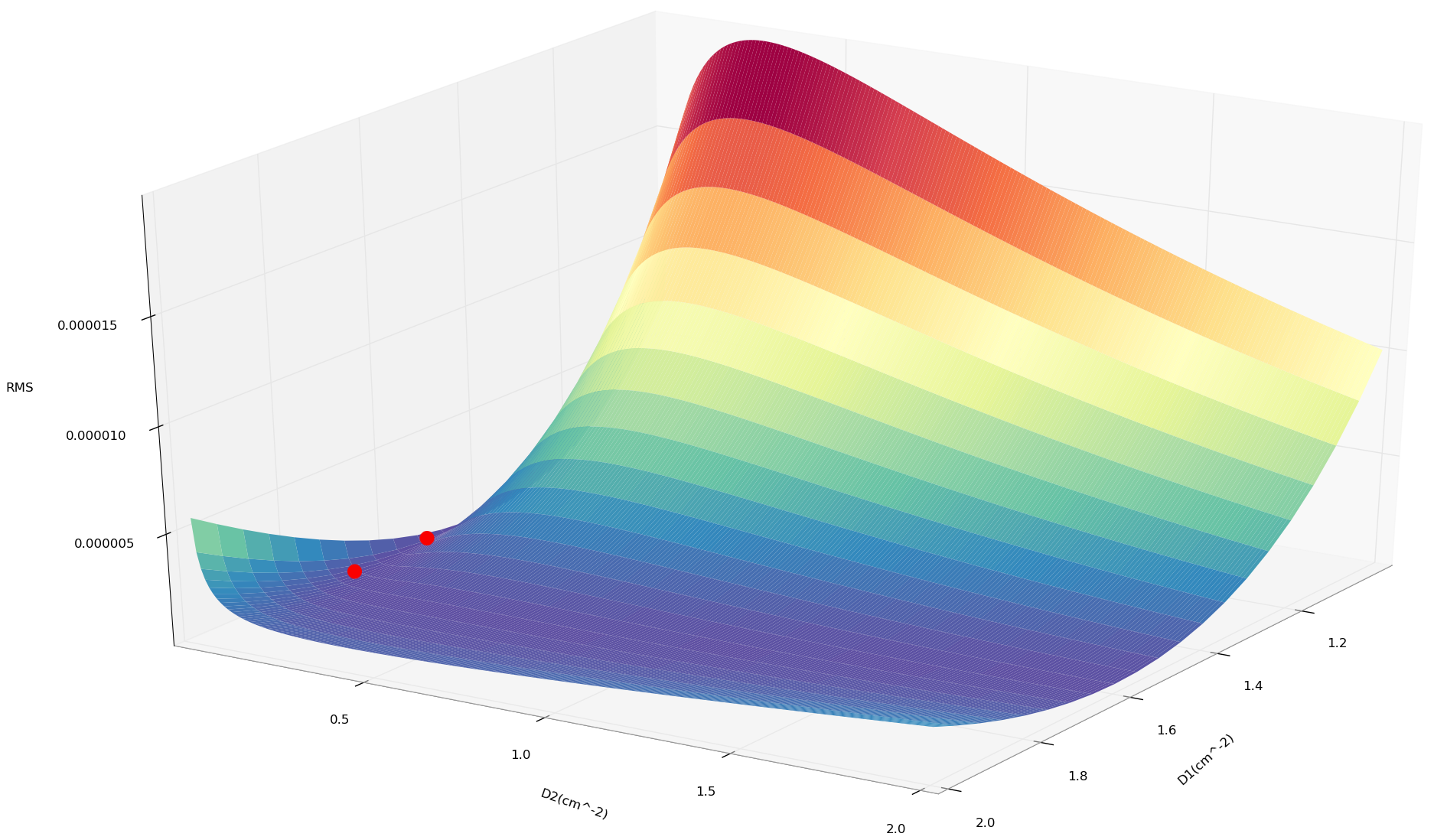}
  \caption{Variations of the functional with ($\text{D}_{\text{1},\text{R}}$,$\text{D}_{\text{2},\text{R}}$)}
  \label{fig:variationsFunc1refl2gp}
 \end{center}
\end{figure}

A radial azimuthal asymmetry is observed because the reflector is homogeneous, and an objective of this study will be to erase this asymmetry. 
\\[1.0em]
Two-parameters optimization are then performed. First a 2-group reflector is computed by optimizing ($\text{D}_{\text{1},\text{R}}$,$\text{D}_{\text{2},\text{R}}$). The variations of the functional with ($\text{D}_{\text{1},\text{R}}$,$\text{D}_{\text{2},\text{R}}$) are presented in Figure~\ref{fig:variationsFunc1refl2gp}.

\begin{figure}[H]
 \begin{center}
  \includegraphics[height=5cm,width=7cm]{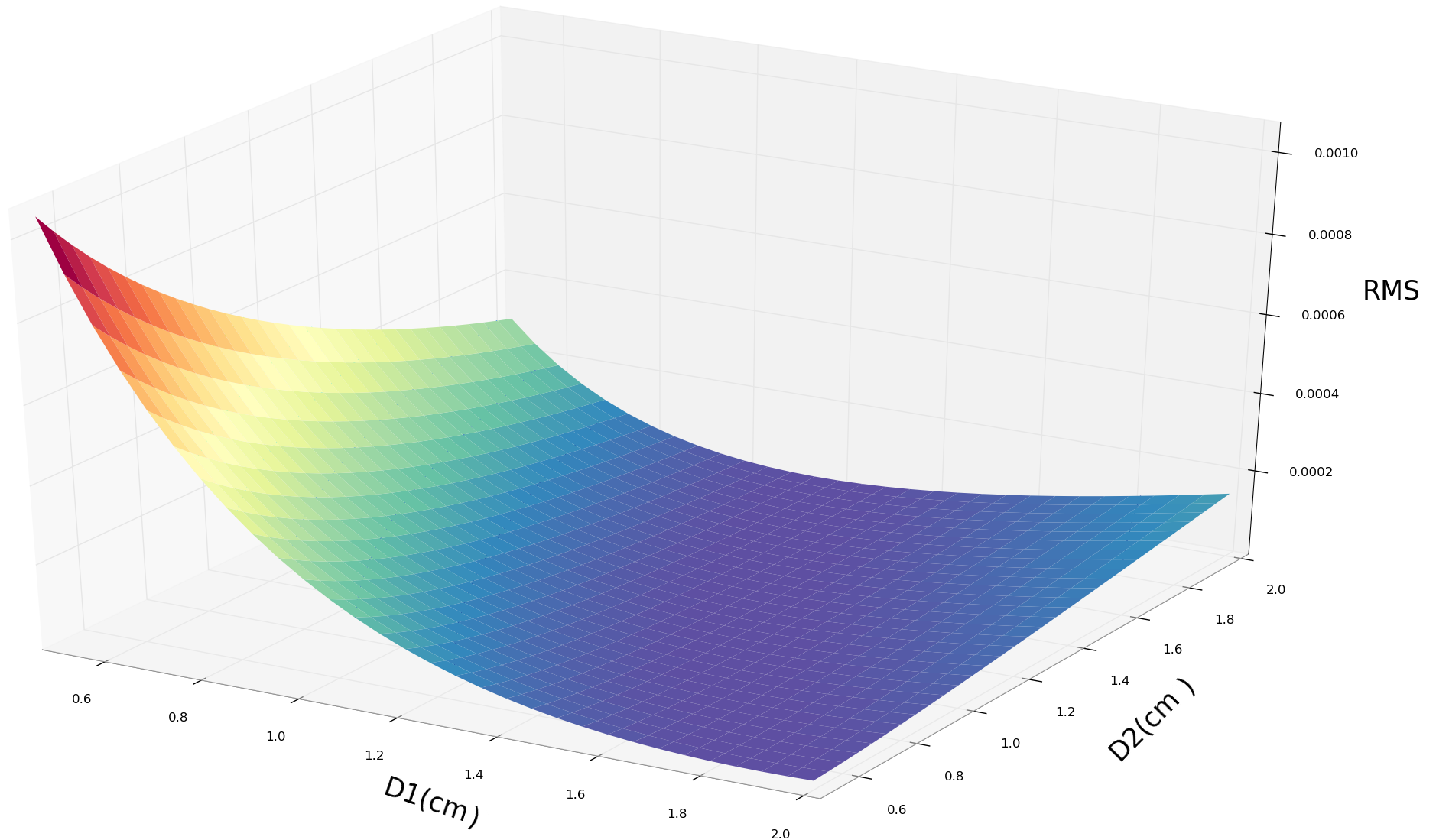}
  \caption{Variations of the functional with ($\text{D}_{\text{1},\text{R}}$,$\text{D}_{\text{2},\text{R}}$)}
  \label{fig:variationsFunc1refl4gpD1D2}
 \end{center}
\end{figure}

\begin{figure}[H]
 \begin{center}
  \includegraphics[height=5cm,width=7cm]{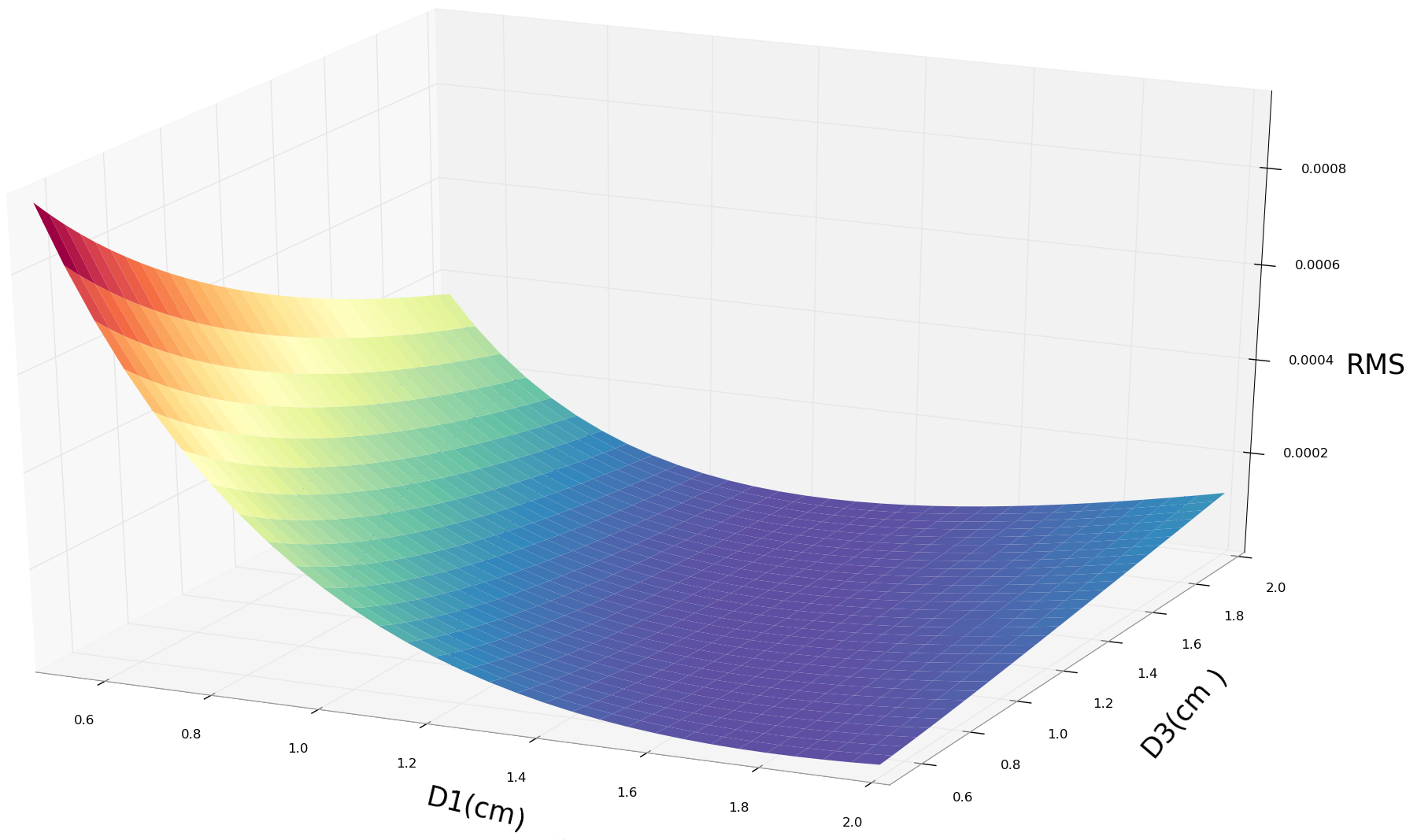}
  \caption{Variations of the functional with ($\text{D}_{\text{1},\text{R}}$,$\text{D}_{\text{3},\text{R}}$)}
  \label{fig:variationsFunc1refl4gpD1D3}
 \end{center}
\end{figure}

The variations of the functional can be very small for large variations of $\text{D}_{\text{1},\text{R}}$ and $\text{D}_{\text{2},\text{R}}$. The parameters are tied, and very different reflectors are obtained with the two computational schemes, with the same background. In fact, the gradients of the functional are calculated according to two different methods in the computational schemes. Given that the parameters are tied, the optimum depends strongly on the initial condition (background and initial gradients).
\\[1.0em]
In order to confirm this observation, 4-group reflectors are computed by optimizing ($\text{D}_{\text{1},\text{R}}$,$\text{D}_{\text{2},\text{R}}$) and ($\text{D}_{\text{1},\text{R}}$,$\text{D}_{\text{3},\text{R}}$). The variations of the functional in this two test cases are presented in Figure~\ref{fig:variationsFunc1refl4gpD1D2} and Figure~\ref{fig:variationsFunc1refl4gpD1D3}.

These two figures confirm the result for 2-group calculations. There is a energy-dependence between the parameters. This doesn't constitute and interesting configuration for our study. In fact, the validation of the computational schemes require an optimization with untied parameters. Moreover, we want to compute a unique optimized reflector.\\[1.0em]
A different test case is then set up. The reflector is divided in two regions as shown on Figure~\ref{fig:figureCoeur2}.

\begin{figure}[H]

 \begin{tiny}
 \tikzstyle{R1}=[rectangle, draw=black, thick, minimum width=0.43cm , minimum height=0.43cm , node distance=0.43cm, text height = 0.5em , fill=gray!10]
 \tikzstyle{R2}=[rectangle, draw=black, thick, minimum width=0.43cm , minimum height=0.43cm , node distance=0.43cm, text height = 0.5em , fill=gray!30]
 \tikzstyle{u1800}=[rectangle, draw=black, thick , minimum width=0.43cm , minimum height=0.43cm , node distance=0.43cm , text height = 0.5em , fill=white!100]
 \tikzstyle{u2412}=[rectangle, draw=black, thick , minimum width=0.43cm , minimum height=0.43cm , node distance=0.43cm , text height = 0.5em , fill=green!30]
 \tikzstyle{u2416}=[rectangle, draw=black, thick , minimum width=0.43cm , minimum height=0.43cm , node distance=0.43cm , text height = 0.5em , fill=cyan!20]
 \tikzstyle{u2420}=[rectangle, draw=black, thick , minimum width=0.43cm , minimum height=0.43cm , node distance=0.43cm , text height = 0.5em , fill=cyan!50]
 \tikzstyle{u3100}=[rectangle, draw=black, thick , minimum width=0.43cm , minimum height=0.43cm , node distance=0.43cm , text height = 0.5em , fill=orange!50]
 \tikzstyle{u3108}=[rectangle, draw=black, thick , minimum width=0.43cm , minimum height=0.43cm , node distance=0.43cm , text height = 0.5em , fill=yellow!30]
 \tikzstyle{u3116}=[rectangle, draw=black, thick , minimum width=0.43cm , minimum height=0.43cm , node distance=0.43cm , text height = 0.5em , fill=red!40]
 \tikzstyle{u3124}=[rectangle, draw=black, thick , minimum width=0.43cm , minimum height=0.43cm , node distance=0.43cm , text height = 0.5em , fill=purple!50]

 \begin{center}
  \begin{tikzpicture}
   \node[name=1,u1800] at (6,0) {1};
   \node[name=2,u2416,right of=1] {2};
   \node[name=3,u1800,right of=2] {3};
   \node[name=4,u2416,right of=3] {4};
   \node[name=5,u1800,right of=4] {5};
   \node[name=6,u3124,right of=5] {6};
   \node[name=7,u1800,right of=6] {7};
   \node[name=8,u2420,right of=7] {8};
   \node[name=9,u3100,right of=8] {9};
   \node[name=34,R1,right of=9] {R1};
   \node[name=35,R1,right of=34] {R1};

   \node[name=10,u1800,above of=2] {10};
   \node[name=11,u2412,right of=10] {11};
   \node[name=12,u1800,right of=11] {12};
   \node[name=13,u2416,right of=12] {13};
   \node[name=14,u1800,right of=13] {14};
   \node[name=15,u2420,right of=14] {15};
   \node[name=16,u3100,right of=15] {16};
   \node[name=17,u3100,right of=16] {17};
   \node[name=36,R1,right of=17] {R1};
   \node[name=37,R1,right of=36] {R1};

   \node[name=18,u1800,above of=11] {18};
   \node[name=19,u2412,right of=18] {19};
   \node[name=20,u1800,right of=19] {20};
   \node[name=21,u2412,right of=20] {21};
   \node[name=22,u1800,right of=21] {22};
   \node[name=23,u3108,right of=22] {23};
   \node[name=38,R1,right of=23] {R1};
   \node[name=39,R1,right of=38] {R1};
   \node[name=40,R1,right of=39] {R1};

   \node[name=24,u1800,above of=19] {24};
   \node[name=25,u2412,right of=24] {25};
   \node[name=26,u1800,right of=25] {26};
   \node[name=27,u3116,right of=26] {27};
   \node[name=28,u3100,right of=27] {28};
   \node[name=41,R1,right of=28] {R1};
   \node[name=42,R1,right of=41] {R1};
   \node[name=43,R1,right of=42] {R1};

   \node[name=29,u1800,above of=25] {29};
   \node[name=30,u2420,right of=29] {30};
   \node[name=31,u3100,right of=30] {31};
   \node[name=44,R2,right of=31] {R2};
   \node[name=45,R2,right of=44] {R2};
   \node[name=46,R2,right of=45] {R2};
   \node[name=47,R2,right of=46] {R2};

   \node[name=32,u3100,above of=30] {32};
   \node[name=33,u3100,right of=32] {33};
   \node[name=48,R2,right of=33] {R2};
   \node[name=49,R2,right of=48] {R2};
   \node[name=50,R2,right of=49] {R2};
   \node[name=51,R2,right of=50] {R2};

   \node[name=52,R2,above of=33] {R2};
   \node[name=53,R2,right of=52] {R2};
   \node[name=54,R2,right of=53] {R2};
   \node[name=55,R2,right of=54] {R2};
   \node[name=56,R2,right of=55] {R2};

   \node[name=57,R2,above of=53] {R2};
   \node[name=58,R2,right of=57] {R2};
   \node[name=59,R2,right of=58] {R2};
   \node[name=60,R2,right of=59] {R2};

   \node[name=61,R2,above of=58] {R2};
   \node[name=62,R2,right of=61] {R2};
   \node[name=63,R2,right of=62] {R2};

   \node[name=64,R2,above of=62] {R2};
   \node[name=65,R2,right of=64] {R2};

   \node[name=66,R2,above of=65] {R2};
  \end{tikzpicture}
 \end{center}
 \end{tiny}
\caption{Eighth of core (1 to 33: core; $\text{R}_{\text{1}}$, $\text{R}_{\text{2}}$: reflector)}
\label{fig:figureCoeur2}
\end{figure}

\begin{figure}[H]
 \begin{center}
  \includegraphics[height=5cm,width=6cm]{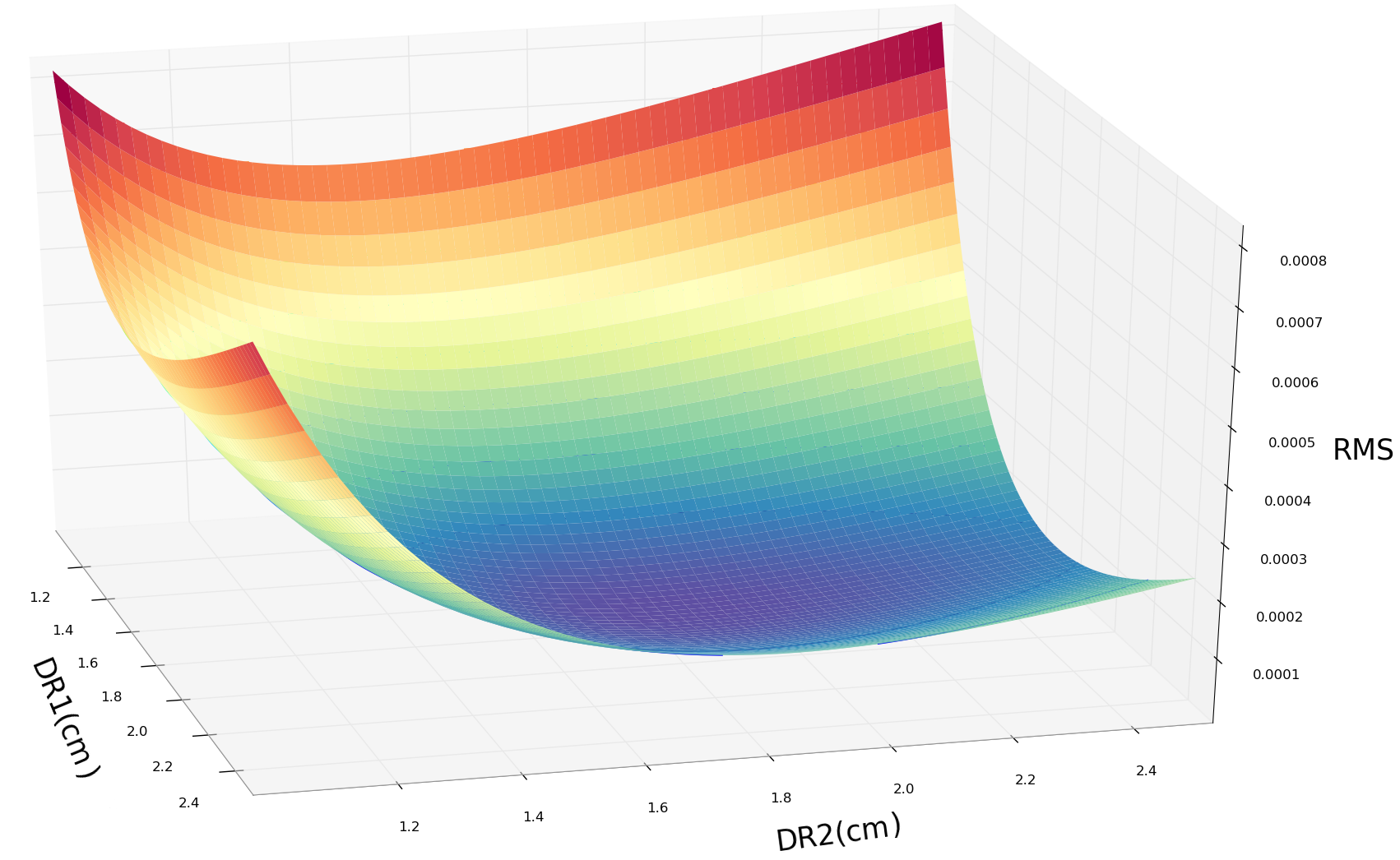}
  \caption{Variations of the functional with ($\text{D}_{\text{1},\text{R}_{\text{1}}};\text{D}_{\text{1},\text{R}_{\text{2}}}$)}
  \label{fig:VarFoncDR1DR2}
 \end{center}
\end{figure}

In this case, a 2-group reflector is computed by optimizing the fast diffusion coefficient in each region of the reflector. The variations of the functional with ($\text{D}_{\text{1},\text{R}_{\text{1}}},\text{D}_{\text{1},\text{R}_{\text{2}}}$) are presented in Figure~\ref{fig:VarFoncDR1DR2}.

\begin{figure}[H]
 \begin{center}
  \includegraphics[scale=0.28]{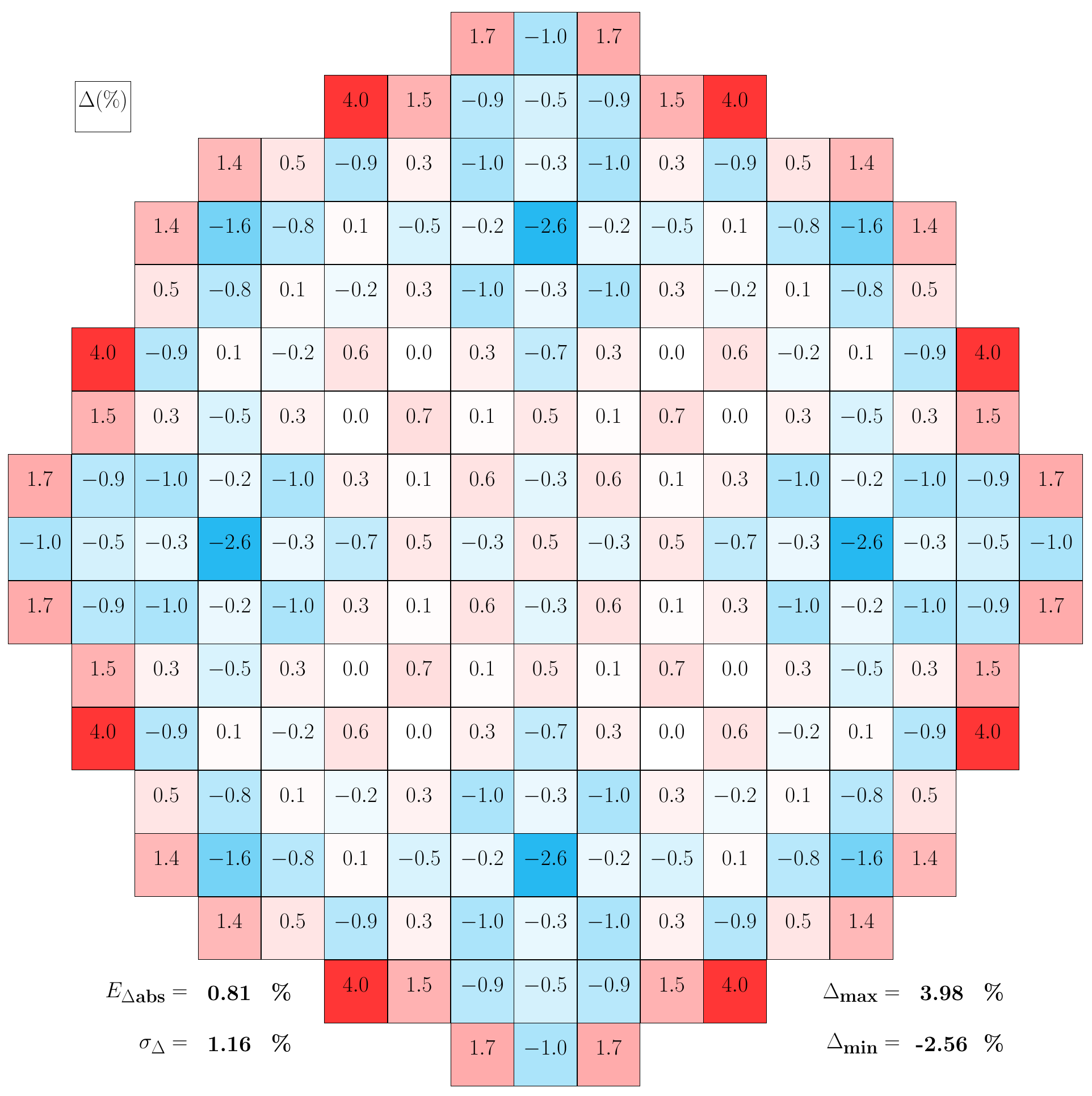}
  \caption{Power discrepancy distribution (validation case 2)}
  \label{fig:puissanceCOCDRAG_D1_2Refl_ADAO}
 \end{center}
\end{figure}

The variations of the functional are very simple in this case, the parameters are untied and the minimum is clearly identifiable. We obtain then very similar results with the OPTEX model and our computational scheme. The discrepancies distribution in this case is presented in Figure~\ref{fig:puissanceCOCDRAG_D1_2Refl_ADAO}.

Our computational scheme have then been validated on several validation test cases. Moreover, a relevant optimization strategy has been identified: reflectors will now be computed by optimizing only the fast group parameter (D or $\Sigma_{\text{tr}}$) in each region of the reflector.

\subsection{Modeling choices}
After this first validation, important modeling choices are made. First, the functional simplified as described in Eq~\ref{eq:Val1} doesn't gather all the information available on the system. In fact, the diffusion operator is less precise in very heterogeneous regions. This is the case of the assemblies at the core/reflector interface and of the assembly 6 (see Figure~\ref{fig:figureCoeur1}), located in a very heterogeneous area. Two classes of assemblies are then identified:\
\begin{enumerate}
  \item[$\bullet$] The class A containing the assemblies in heterogeneous regions: the errors on the observations made for this class are higher: $\sigma_{\text{R,j}} = 5, \: \forall j \in A=\{\text{6,9,17,23,28,31,33}\}$.\
  \item[$\bullet$] The class B containing the other assemblies: the errors on the observations made for this class are less important: $\sigma_{\text{R,j}} = 0.5, \: \forall j \notin A$.\
\end{enumerate}
The functional for diffusion calculations is then:

\begin{equation}
J(\textbf{x}) = \sum_{j\in B} \left(\frac{P_{j}(\textbf{x})}{\sum_{k=1}^{33}P_{k}(\textbf{x})}-\frac{P^{*}_{j}}{\sum_{k=1}^{33} P^{*}_{k}}\right)^{2}
 + \sum_{j\in A} \frac{1}{10}\left(\frac{P_{j}(\textbf{x})}{\sum_{k=1}^{33}P_{k}(\textbf{x})}-\frac{P^{*}_{j}}{\sum_{k=1}^{33} P^{*}_{k}}\right)^{2} 
\label{eq:foncAssDonnDiff}
\end{equation}

\begin{figure}[H]

 \begin{tiny}
 \tikzstyle{R1}=[rectangle, draw=black, thick, minimum width=0.43cm , minimum height=0.43cm , node distance=0.43cm, text height = 0.5em , fill=gray!10]
 \tikzstyle{R2}=[rectangle, draw=black,thick, minimum width=0.43cm , minimum height=0.43cm , node distance=0.43cm, text height = 0.5em , fill=gray!30]
  \tikzstyle{R3}=[rectangle, draw=black,thick, minimum width=0.43cm , minimum height=0.43cm , node distance=0.43cm, text height = 0.5em , fill=gray!70]
 \tikzstyle{u1800}=[rectangle, draw=black, thick , minimum width=0.43cm , minimum height=0.43cm , node distance=0.43cm , text height = 0.5em , fill=white!100]
 \tikzstyle{u2412}=[rectangle, draw=black, thick , minimum width=0.43cm , minimum height=0.43cm , node distance=0.43cm , text height = 0.5em , fill=green!30]
 \tikzstyle{u2416}=[rectangle, draw=black, thick , minimum width=0.43cm , minimum height=0.43cm , node distance=0.43cm , text height = 0.5em , fill=cyan!20]
 \tikzstyle{u2420}=[rectangle, draw=black, thick , minimum width=0.43cm , minimum height=0.43cm , node distance=0.43cm , text height = 0.5em , fill=cyan!50]
 \tikzstyle{u3100}=[rectangle, draw=black, thick , minimum width=0.43cm , minimum height=0.43cm , node distance=0.43cm , text height = 0.5em , fill=orange!50]
 \tikzstyle{u3108}=[rectangle, draw=black, thick , minimum width=0.43cm , minimum height=0.43cm , node distance=0.43cm , text height = 0.5em , fill=yellow!30]
 \tikzstyle{u3116}=[rectangle, draw=black, thick , minimum width=0.43cm , minimum height=0.43cm , node distance=0.43cm , text height = 0.5em , fill=red!40]
 \tikzstyle{u3124}=[rectangle, draw=black, thick , minimum width=0.43cm , minimum height=0.43cm , node distance=0.43cm , text height = 0.5em , fill=purple!50]

 \begin{center}
  \begin{tikzpicture}
   \node[name=1,u1800] at (6,0) {1};
   \node[name=2,u2416,right of=1] {2};
   \node[name=3,u1800,right of=2] {3};
   \node[name=4,u2416,right of=3] {4};
   \node[name=5,u1800,right of=4] {5};
   \node[name=6,u3124,right of=5] {6};
   \node[name=7,u1800,right of=6] {7};
   \node[name=8,u2420,right of=7] {8};
   \node[name=9,u3100,right of=8] {9};
   \node[name=34,R1,right of=9] {R1};
   \node[name=35,R1,right of=34] {R1};

   \node[name=10,u1800,above of=2] {10};
   \node[name=11,u2412,right of=10] {11};
   \node[name=12,u1800,right of=11] {12};
   \node[name=13,u2416,right of=12] {13};
   \node[name=14,u1800,right of=13] {14};
   \node[name=15,u2420,right of=14] {15};
   \node[name=16,u3100,right of=15] {16};
   \node[name=17,u3100,right of=16] {17};
   \node[name=36,R1,right of=17] {R1};
   \node[name=37,R1,right of=36] {R1};

   \node[name=18,u1800,above of=11] {18};
   \node[name=19,u2412,right of=18] {19};
   \node[name=20,u1800,right of=19] {20};
   \node[name=21,u2412,right of=20] {21};
   \node[name=22,u1800,right of=21] {22};
   \node[name=23,u3108,right of=22] {23};
   \node[name=38,R3,right of=23] {R3};
   \node[name=39,R1,right of=38] {R1};
   \node[name=40,R1,right of=39] {R1};

   \node[name=24,u1800,above of=19] {24};
   \node[name=25,u2412,right of=24] {25};
   \node[name=26,u1800,right of=25] {26};
   \node[name=27,u3116,right of=26] {27};
   \node[name=28,u3100,right of=27] {28};
   \node[name=41,R3,right of=28] {R4};
   \node[name=42,R1,right of=41] {R1};
   \node[name=43,R1,right of=42] {R1};

   \node[name=29,u1800,above of=25] {29};
   \node[name=30,u2420,right of=29] {30};
   \node[name=31,u3100,right of=30] {31};
   \node[name=44,R3,right of=31] {R5};
   \node[name=45,R2,right of=44] {R2};
   \node[name=46,R2,right of=45] {R2};
   \node[name=47,R2,right of=46] {R2};

   \node[name=32,u3100,above of=30] {32};
   \node[name=33,u3100,right of=32] {33};
   \node[name=48,R3,right of=33] {R6};
   \node[name=49,R2,right of=48] {R2};
   \node[name=50,R2,right of=49] {R2};
   \node[name=51,R2,right of=50] {R2};

   \node[name=52,R2,above of=33] {R2};
   \node[name=53,R2,right of=52] {R2};
   \node[name=54,R2,right of=53] {R2};
   \node[name=55,R2,right of=54] {R2};
   \node[name=56,R2,right of=55] {R2};

   \node[name=57,R2,above of=53] {R2};
   \node[name=58,R2,right of=57] {R2};
   \node[name=59,R2,right of=58] {R2};
   \node[name=60,R2,right of=59] {R2};

   \node[name=61,R2,above of=58] {R2};
   \node[name=62,R2,right of=61] {R2};
   \node[name=63,R2,right of=62] {R2};

   \node[name=64,R2,above of=62] {R2};
   \node[name=65,R2,right of=64] {R2};

   \node[name=66,R2,above of=65] {R2};
  \end{tikzpicture}
 \end{center}
 \end{tiny}
\caption{Eighth of core (1 to 33: core; $\{\text{R}_{\text{1}}$, $\text{R}_{\text{6}}\}$: reflector)}
\label{fig:figureCoeur3}
\end{figure}
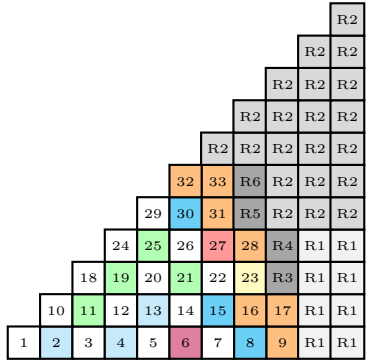

When $\text{SP}_{\text{N}}$ calculations are performed, the assemblies in the active core are modeled pin by pin, and no class distinction can be made between the assemblies. In this case, the functional of Eq~\ref{eq:foncAssDonn3} will be used.
\\[1.0em]
The most important power discrepancies are observed at the core/reflector interface. To reduce these discrepancies, the reflector meshing of Figure~\ref{fig:figureCoeur3} is proposed. 

This meshing takes into account the reflector's geometry. Only six-parameter optimizations will then be performed: \textbf{x}=\textbf{D}=($\text{D}_{\text{1},\text{R}_{\text{i}}},\: i \in \text{[1,6]})$ for diffusion calculations and \textbf{x}=$\boldsymbol\Sigma_{\textbf{tr}}$=$(\Sigma_{\text{tr,1},\text{R}_{\text{i}}},\: i \in \text{[1,6]})$ for $\text{SP}_{\text{N}}$ calculations.

\section{Results for diffusion calculations}

Here, the results of the computational scheme are presented for diffusion calculations, with the modeling choices of Section 2.3. The results for diffusion calculations have been obtained by Clerc \citep{refMoi}.

\subsection{2-group calculations}
\begin{figure}[H]
 \begin{center}
  \includegraphics[scale=0.29]{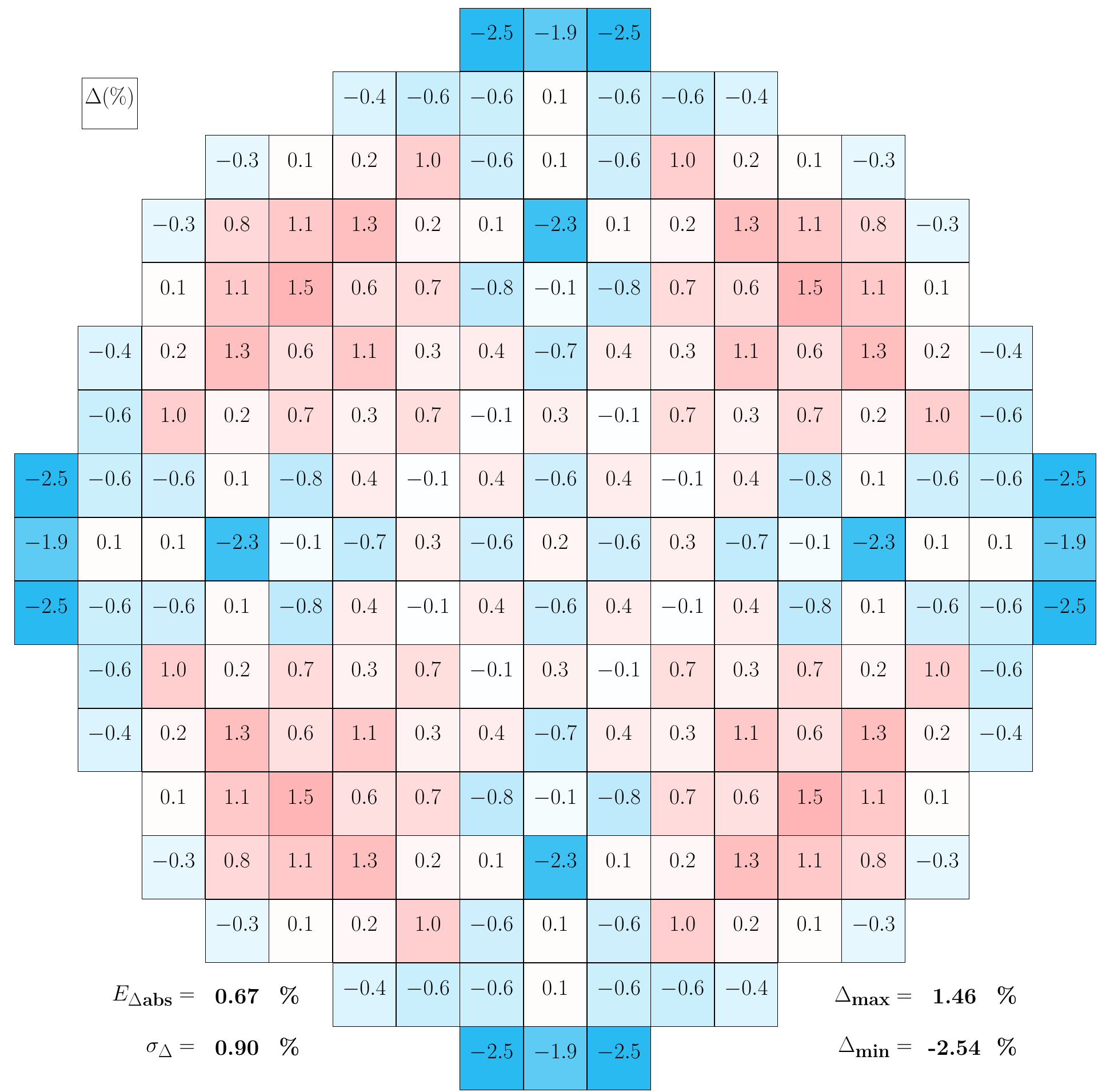}
  \caption{Power discrepancy distribution (before opt.)}
  \label{fig:puissanceCOCDRAG_D11D16_LefLeb_AvantOpt}
 \end{center}
\end{figure}

\begin{figure}[H]
 \begin{center}
  \includegraphics[scale=0.29]{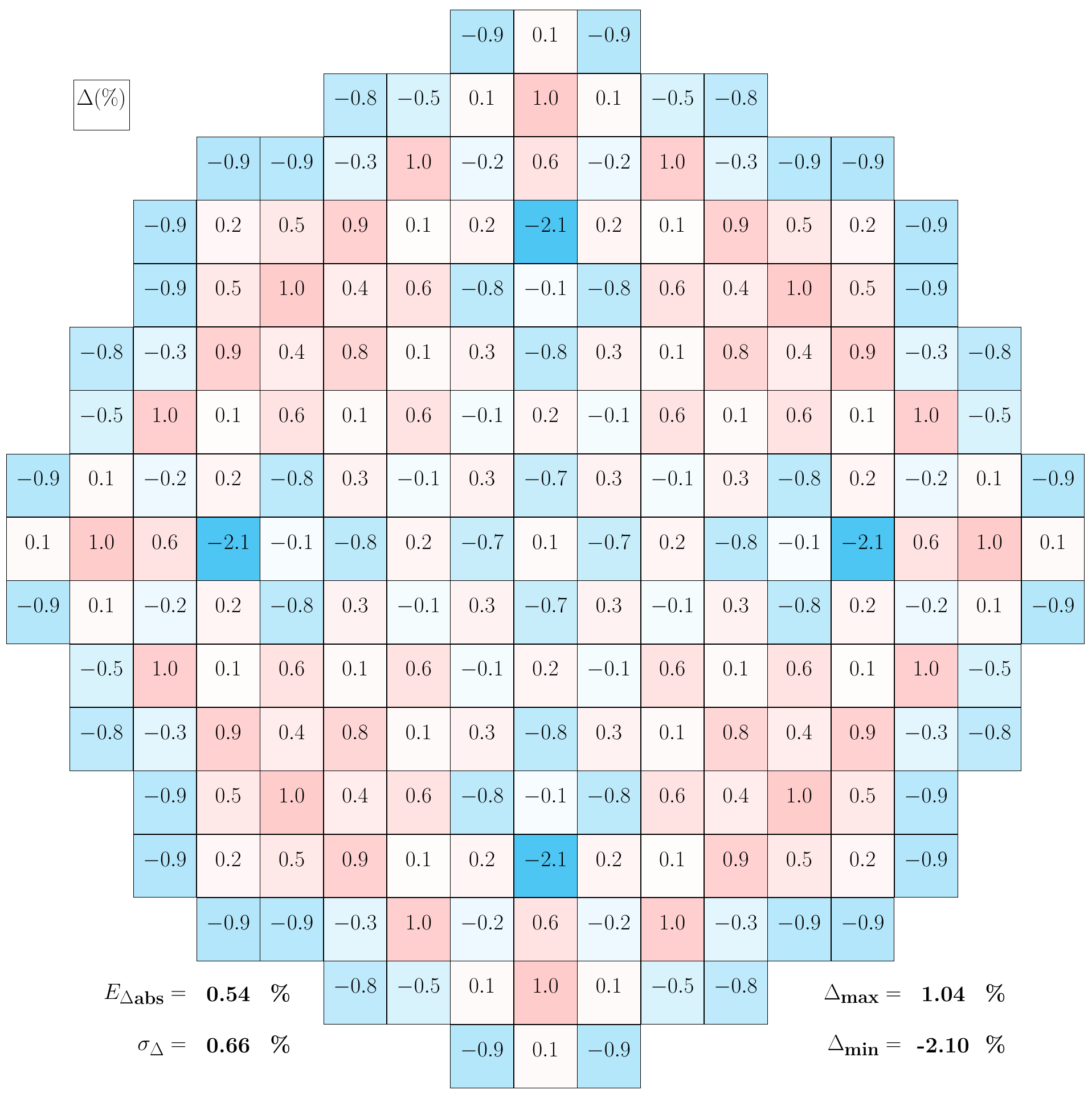}
  \caption{Power discrepancy distribution (after opt.)}
  \label{fig:puissanceCOCDRAG_D11D16_LefLeb_ApresOpt}
 \end{center}
\end{figure}

The reflector is initialized as explained in Section 2.4. The discrepancies distributions before and after optimization are presented in Figure~\ref{fig:puissanceCOCDRAG_D11D16_LefLeb_AvantOpt} and Figure~\ref{fig:puissanceCOCDRAG_D11D16_LefLeb_ApresOpt}.

\begin{figure}[H]
 \begin{center}
  \includegraphics[scale=0.1]{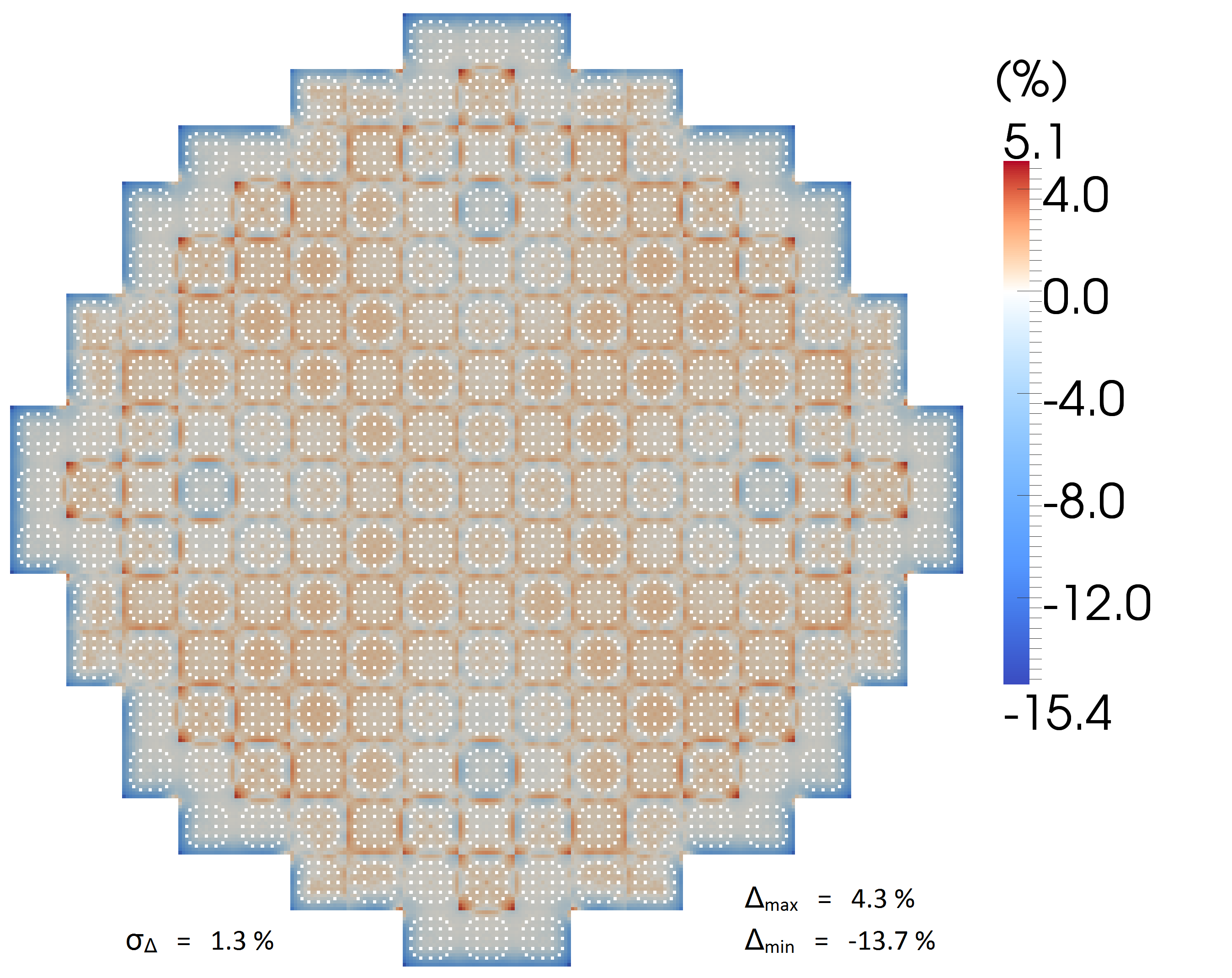}
  \caption{Power discrepancy distribution pin by pin after optimization}
  \label{fig:RFPApresOpt-2gr}
 \end{center}
\end{figure}

The initial Lefebvre-Lebigot reflector is well designed, but a slight effect of our computational scheme is observed (\mbox{$\text{E}_{\Delta_{\text{abs}}}$=0.67 \%} and \mbox{$\sigma_{\Delta}$=0.9 \%} initially and respectively \mbox{0.54 \%} and \mbox{0.66 \%} after optimization). Moreover, we observe a significant improvement of the azimuthal asymmetry observed in Figure~\ref{fig:puissanceCOCDRAG_D11D16_LefLeb_AvantOpt}: initially, the discrepancy is \mbox{-2.5 \%} in assembly 17 and \mbox{1.5 \%} in assembly 29, and after optimization the discrepancy is \mbox{-0.9 \%} in assembly 17 and \mbox{1.0 \%} in assembly 29. In fact, the initial reflector is homogeneous and we compute a 6 regions reflector that describes better the heterogeneities of the reflector's structure.
\\[1.0em]
A pin-power reconstruction is then performed on the power distribution after optimization, in order to obtain the pin-by-pin power discrepancies distribution presented on Figure~\ref{fig:RFPApresOpt-2gr}.

Here we note that the distribution is very homogeneous, with higher discrepancies located in the last rows of pins at the interface core/reflector.
The two reflectors have also been compared in terms of critical boron with inserted rods discrepancies, and rod banks efficiencies discrepancies with experimental data, before and after optimization (respectively $\Delta^{\text{(i)}}_{\text{Cb}}$, $\Delta^{\text{(f)}}_{\text{Cb}}$, $\Delta^{\text{(i)}}_{\text{eff}}$ and $\Delta^{\text{(f)}}_{\text{eff}}$ in \%) in Table~\ref{tab:tab2gpDiff}, computed as described in \citep{refMaitrise}. To obtain these data, 3D calculations have been performed. The upper and lower reflector are obtained with the Lefebvre-Lebigot method, but they are not optimized as our computational scheme only computes 2D radial equivalent reflectors.

\begin{tiny}
\begin{table}[H]
 \caption{Critical boron and rod banks efficiencies}
 \centering
 \begin{tabular}{|c||c|c||c|c|}
  \hline
  Banks & $\Delta^{\text{(i)}}_{\text{Cb}}$ & $\Delta^{\text{(f)}}_{\text{Cb}}$ & $\Delta^{\text{(i)}}_{\text{eff}}$ & $\Delta^{\text{(f)}}_{\text{eff}}$ \\
  \hline
  \hline
  A & 0.2 & 0.2 & 2.4 & 2.4\\
  \hline
  B & 0.0 & 0.1 & -2.3 & -2.5\\
  \hline
  C & -0.2 & -0.3 & -5.5 & -5.5\\
  \hline
  D & 0.0 & 0.1 & 2.0 & 2.0\\
  \hline
  E & -1.1 & -1.1 & 2.8 & 2.8\\
  \hline
  F & -1.3 & -1.2 & -0.1 & 0.0\\
  \hline
  G & -0.9 & 0.0 & -3.8 & -4.4\\
  \hline
  H & -7.2 & -7.3 & -0.2 & 0.0\\
  \hline
 \end{tabular}
 \label{tab:tab2gpDiff}
\end{table} 
\end{tiny}

The initial reflector is well designed, and no significant improvement on the critical boron and on the rods banks efficiencies are observed when using the reflector obtained with our computational scheme. Even a slight deterioration of the results is observed, for rod bank D (discrepancy on critical boron from \mbox{0.0 \%} to \mbox{0.1 \%}) or rod bank G (discrepancy on integral efficiency from \mbox{-3.8 \%} to \mbox{-4.4 \%}). However, these variations are very small compared to the experimental values, and they are not significant. The two reflectors, before and after optimization, are too close to show any improvement on more macroscopic indicators such as critical boron or rods woth.

\subsection{4-group calculations}
\begin{figure}[H]
 \begin{center}
  \includegraphics[scale=0.28]{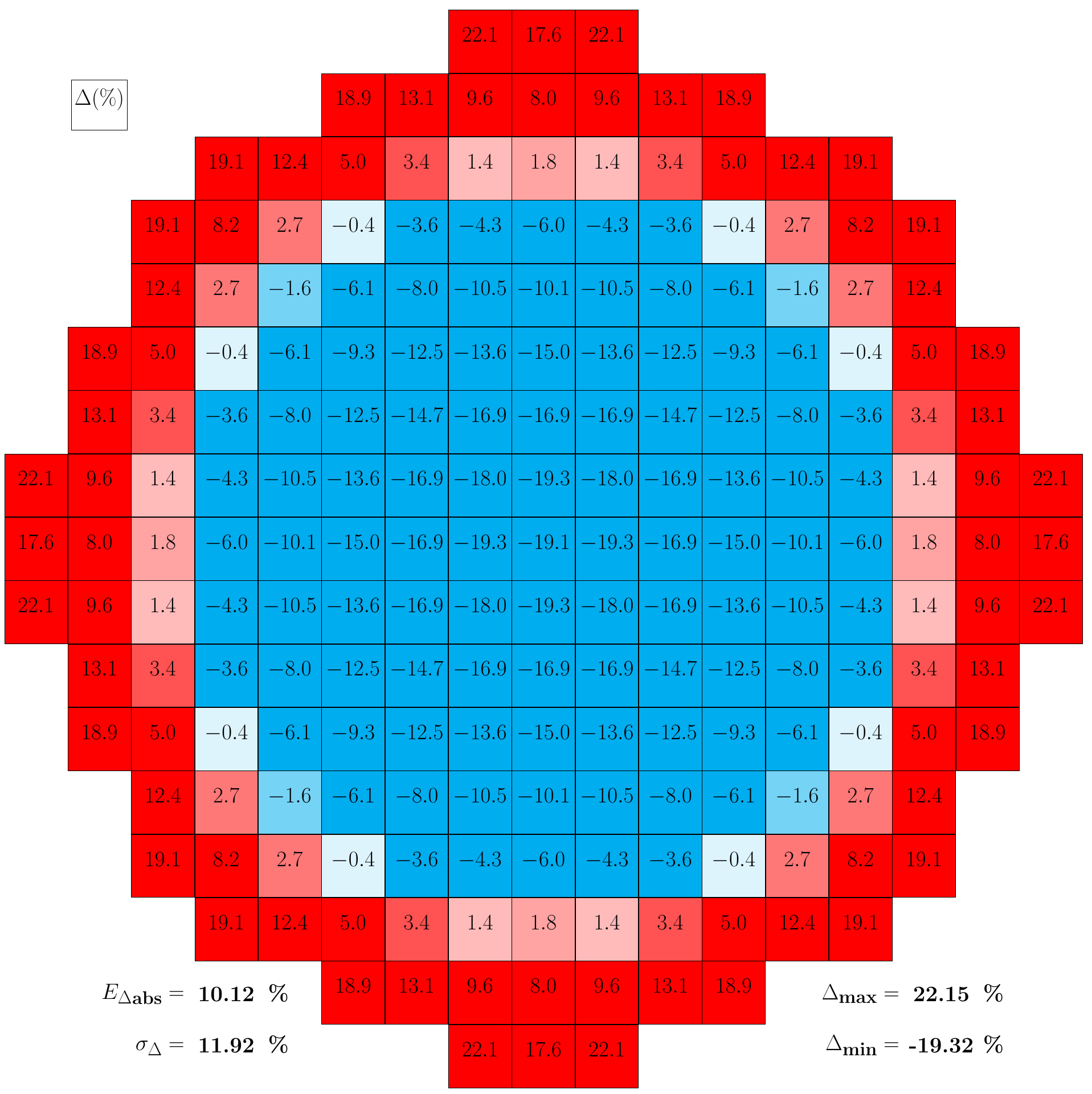}
  \caption{Power discrepancy distribution (before opt.)}
  \label{fig:puissanceCOCDRAG_4grDiff_D11D16_AvantOpt}
 \end{center}
\end{figure}

\begin{figure}[H]
 \begin{center}
  \includegraphics[scale=0.28]{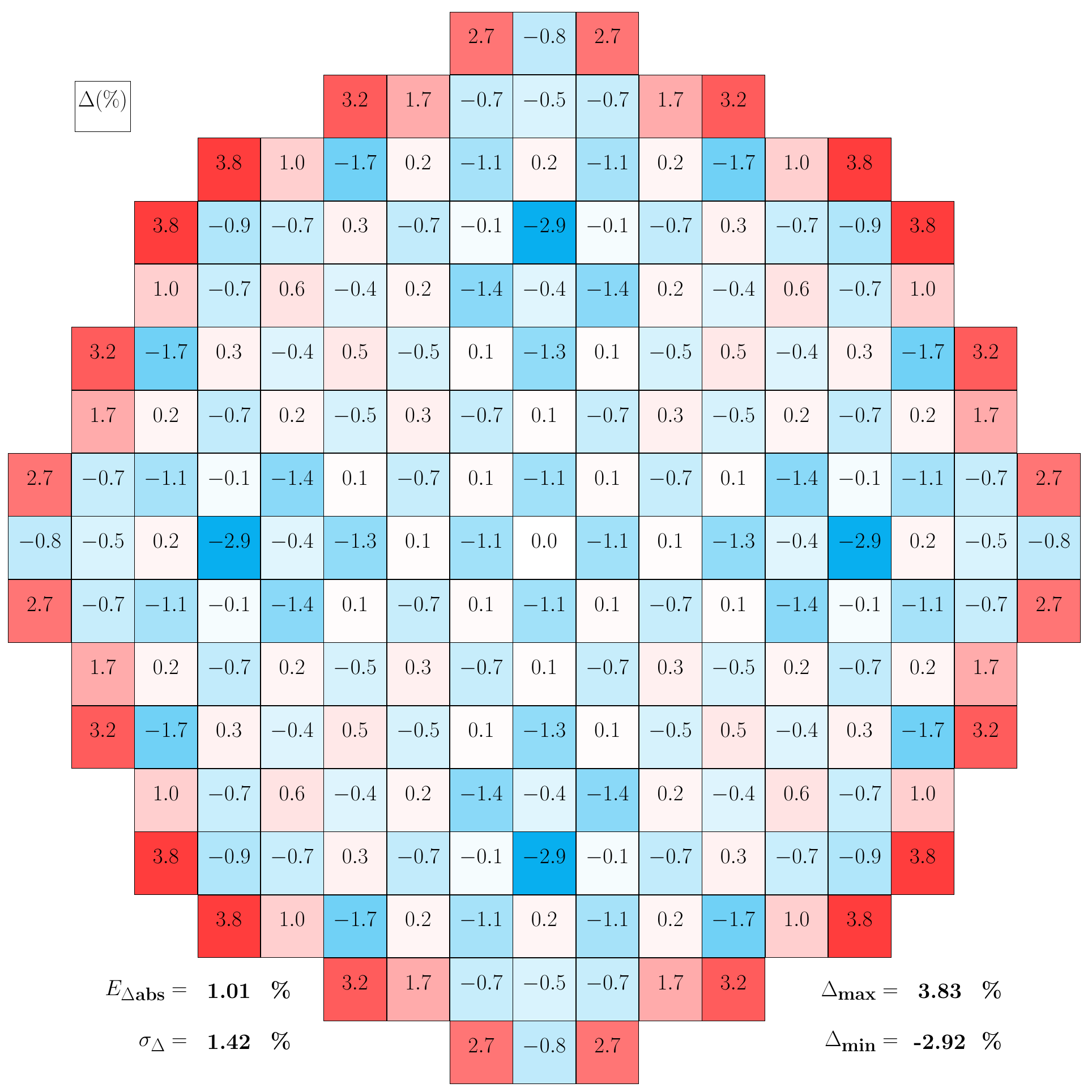}
  \caption{Power discrepancy distribution (after opt.)}
  \label{fig:puissanceCOCDRAG_4grDiff_D11D16_ADAO}
 \end{center}
\end{figure}

Here, 4-group calculations are performed. The discrepancies distributions between the power computed by COCAGNE and the reference power before and after optimization are presented respectively in Figure~\ref{fig:puissanceCOCDRAG_4grDiff_D11D16_AvantOpt} and Figure~\ref{fig:puissanceCOCDRAG_4grDiff_D11D16_ADAO}. 

\begin{figure}[H]
 \begin{center}
  \includegraphics[scale=0.1]{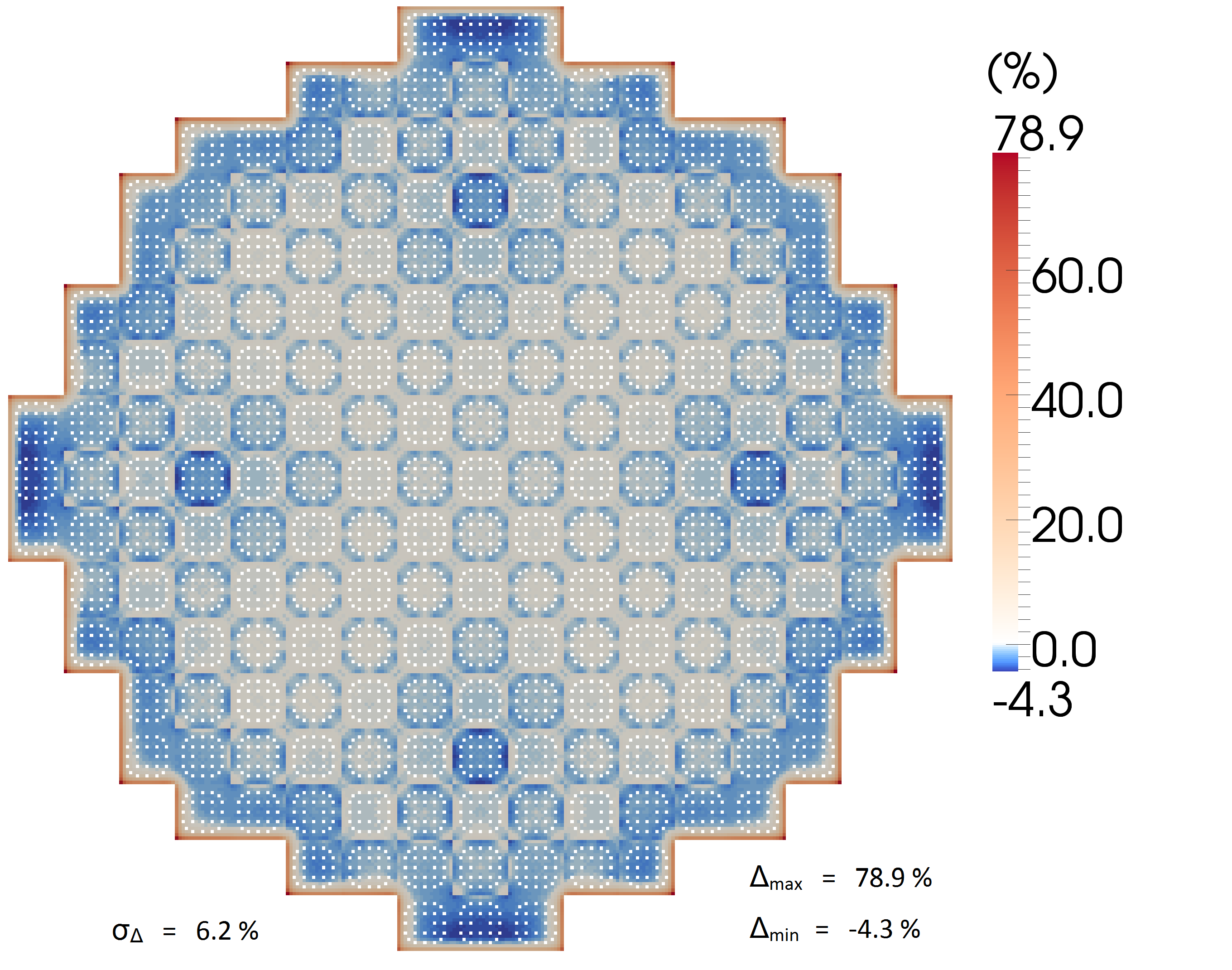}
  \caption{Power discrepancy distribution pin by pin after optimization}
  \label{fig:RFPApresOpt-4gr}
 \end{center}
\end{figure}

Here, the initial reflector is poorly designed, as shown in Section 3.1. The diffusion coefficients are recovered from the reference calculation ($\text{D}=\int_{E}\frac{\text{dE}}{3\Sigma_{\text{tr}}\text{(E)}}$), and this approximation is not physically relevant. Figure~\ref{fig:puissanceCOCDRAG_4grDiff_D11D16_AvantOpt} then shows very important discrepancies, with a very strong radial asymmetry (\mbox{$\Delta$ = -19.1 \%} in assembly 1 and \mbox{17.6 \%} in assembly 9). 
\\[1.0em]
We observe a significant improvement of the discrepancy distribution after optimization: \mbox{$\text{E}_{\Delta_{\text{abs}}}$=10.12 \%} and \mbox{$\sigma_{\Delta}$=11.92 \%} initially and respectively \mbox{1.01 \%} and \mbox{1.42 \%} after optimization. Though, the reflector obtained here is less performant than the Lefebvre-Lebigot reflector. 
\\[1.0em]
The pin-power reconstruction is also performed here, and the pin-by-pin discrepancies distribution is presented in Figure~\ref{fig:RFPApresOpt-4gr}. This distribution shows very high discrepancies in the last rows of pins at the core/reflector interface (\mbox{$\Delta_{\text{max}}$ = 78.9 \%}). These very high discrepancies are the main limit of this study. In fact, they are unacceptable in full core simulations (for vessel fluence calculation for instance), and several possible solutions to this issue will be proposed in the discussion.

The reflectors are then compared in terms of critical boron discrepancies and rod banks efficiencies discrepancies with experimental data, before and after optimization in Table~\ref{tab:tab4gpDiff}, as described in Section 4.1. Here, as the Lefebvre-Lebigot method can't be used, the upper and lower reflector are equal to the optimized radial reflector, homogenized to one region.

\begin{tiny}
\begin{table}[H]
 \caption{Critical boron and rod banks efficiencies}
 \centering
 \begin{tabular}{|c||c|c||c|c|}
  \hline
  Banks & $\Delta^{\text{(i)}}_{\text{Cb}}$ & $\Delta^{\text{(f)}}_{\text{Cb}}$ & $\Delta^{\text{(i)}}_{\text{eff}}$ & $\Delta^{\text{(f)}}_{\text{eff}}$ \\
  \hline
  \hline
  A & 2.4 & 0.2 & 1.6 & 2.4\\
  \hline
  B & 2.5 & 0.1 & -2.5 & -2.1\\
  \hline
  C & 2.7 & -0.3 & -5.5 & -5.5\\
  \hline
  D & 4.2 & 0.1 & 0.8 & 1.9\\
  \hline
  D & 3.3 & -1.0 & 3.4 & 2.7\\
  \hline
  F & 2.2 & -1.2 & 1.9 & 0.1\\
  \hline
  G & 8.8 & -0.4 & -9.3 & -4.3\\
  \hline
  H & -3.4 & -6.6 & 2.3 & 0.0\\
  \hline
 \end{tabular}
 \label{tab:tab4gpDiff}
\end{table} 
\end{tiny}

Here again, the results are significantly improved. However, a deterioration is observed for rod bank H (\mbox{$\Delta^{\text{(i)}}_{\text{Cb}}$ = -3.4 \%} before optimization and \mbox{$\Delta^{\text{(f)}}_{\text{Cb}}$ = -6.6 \%}), and D (\mbox{$\Delta^{\text{(i)}}_{\text{eff}}$ = 0.8 \%} before optimization and \mbox{$\Delta^{\text{(f)}}_{\text{eff}}$ = 1.9 \%}). The results finally deteriorated are very close to the ones observed in Table~\ref{tab:tab2gpDiff} obtained with the Lefebvre-Lebigot reflector. They enable an overall improvement of the results.
It is interesting to confirm that our computational scheme is able to compute a good 4-group reflector with a very approximate initial reflector.

\subsection{Conclusion on diffusion calculations}
The results presented in this section show first that for 2-group calculation, the initial Lefebvre-Lebigot reflector is very well designed and the effect on our computational scheme is not significant. Though, the azimuthal asymmetry shown with the initial reflector is well erased by the optimization because the optimized reflector is modeled using 6 different regions. Secondly, the results show that for 4-group calculation, even if the initial reflector leads to very bad results, our computational scheme has a strong effect on reducing the power discrepancies. Though, the results are less satisfying than the optimized Lefebvre-Lebigot reflector. In an industrial perspective, 2-group calculations are much faster than 4-group calculation. With equal or very similar precision of the two reflectors, the 2-group reflector would then be preferred to the 4-group reflector.

\section{Results for $\text{SP}_{\text{3}}$ calculations}

Here are presented the results of our computational scheme for $\text{SP}_{\text{3}}$ calculations. The modeling choices in this case are presented in Section 3.3. The functional minimized by our computational scheme in this section is presented in Eq~\ref{eq:foncAssDonn3}, and the reflector is divided into 6 zones presented in Figure~\ref{fig:figureCoeur3}. 

\subsection{Results for 4-group calculations}
\begin{figure}[H]
 \begin{center}
  \includegraphics[scale=0.28]{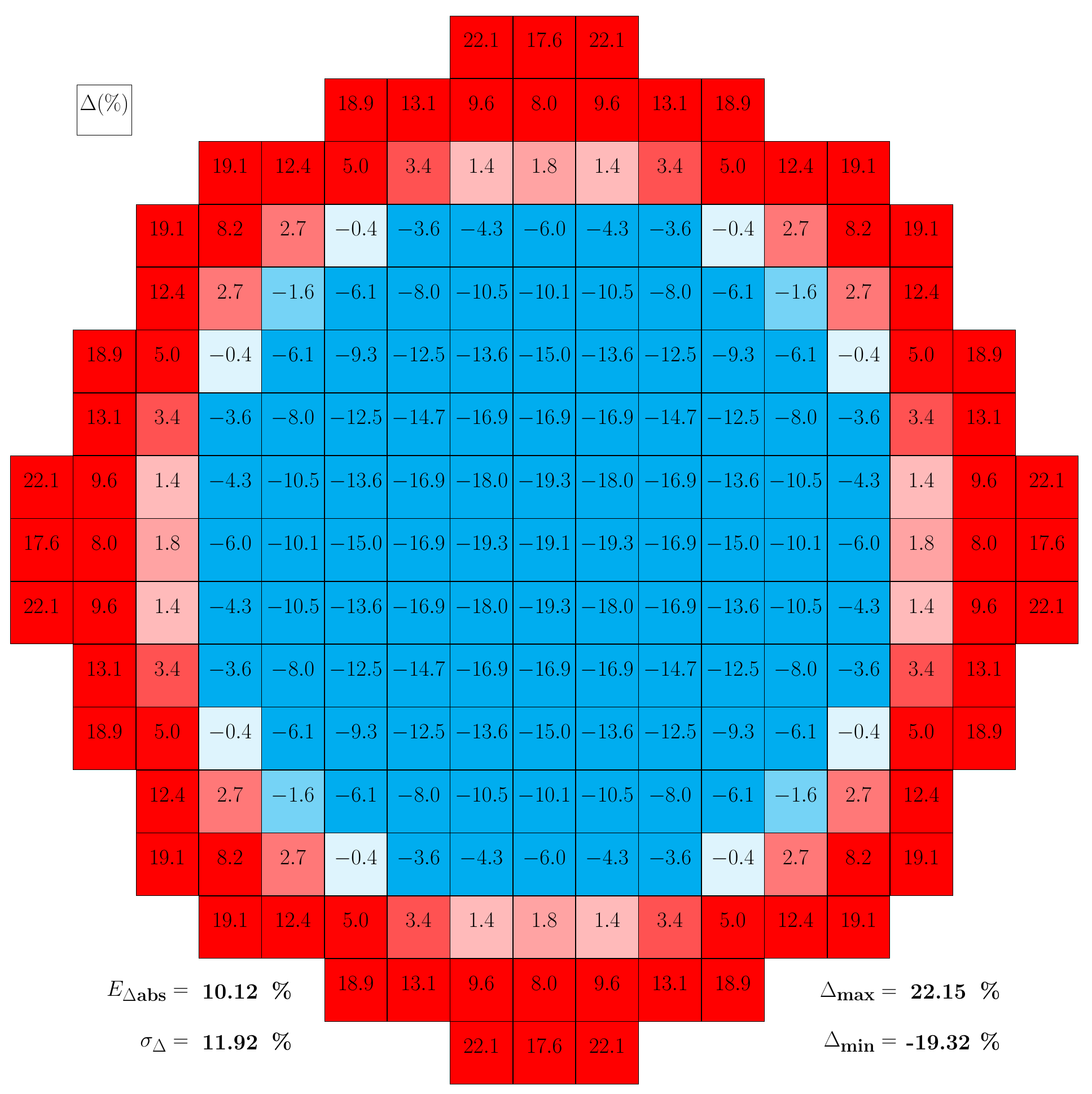}
  \caption{Power discrepancy distribution (before opt.)}
  \label{fig:puissanceCOCDRAG_4grSPN_D11D16_AvantOpt}
 \end{center}
\end{figure}

\begin{figure}[H]
 \begin{center}
  \includegraphics[scale=0.28]{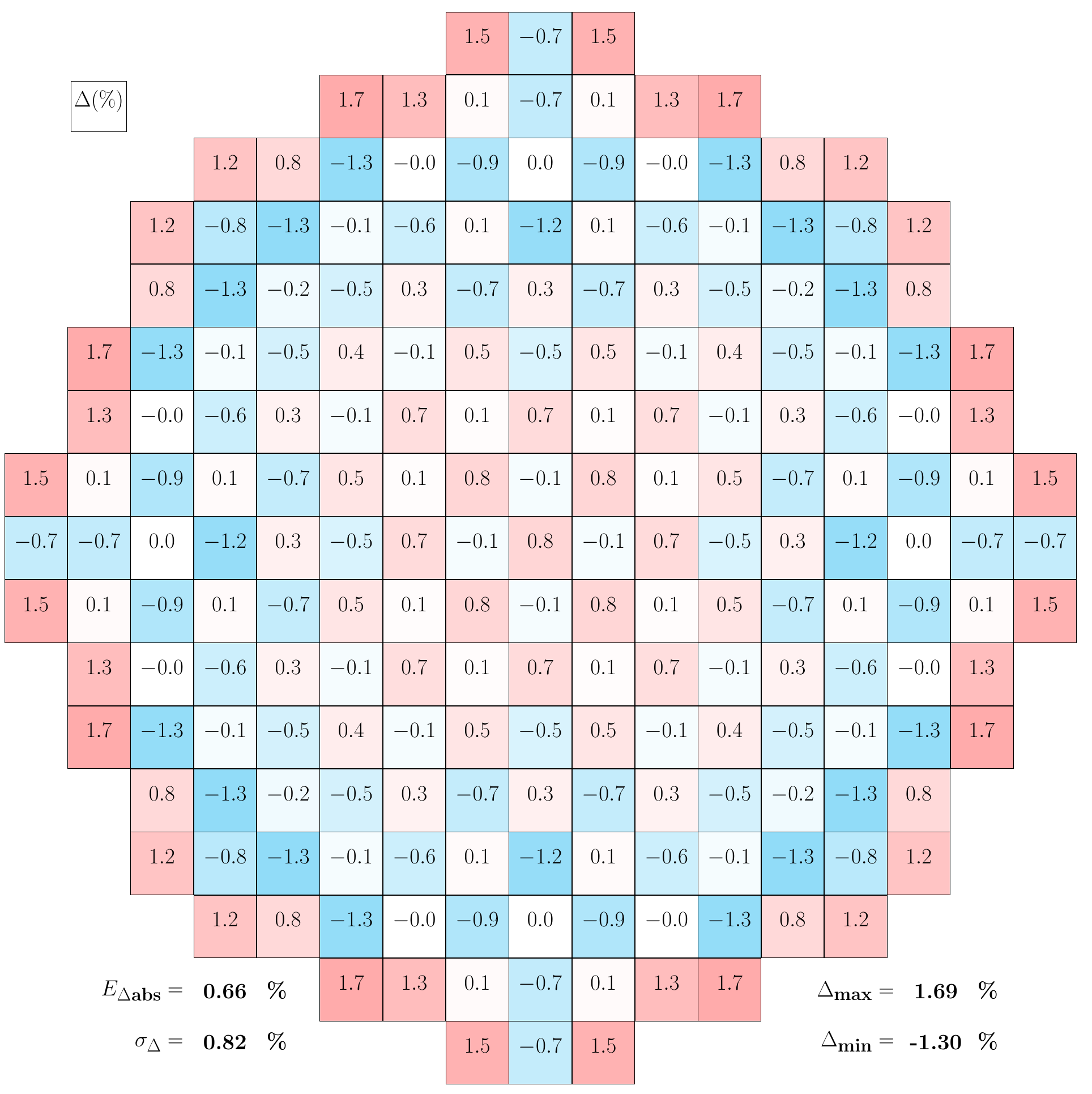}
  \caption{Power discrepancy distribution (after opt.)}
  \label{fig:puissanceCOCDRAG_4grSPN_D11D16_ADAO}
 \end{center}
\end{figure}

In this Section, the results for 4-group calculations are presented. The discrepancies distributions before and after optimization are presented in Figure~\ref{fig:puissanceCOCDRAG_4grSPN_D11D16_AvantOpt} and Figure~\ref{fig:puissanceCOCDRAG_4grSPN_D11D16_ADAO}.

\begin{figure}[H]
 \begin{center}
  \includegraphics[scale=0.1]{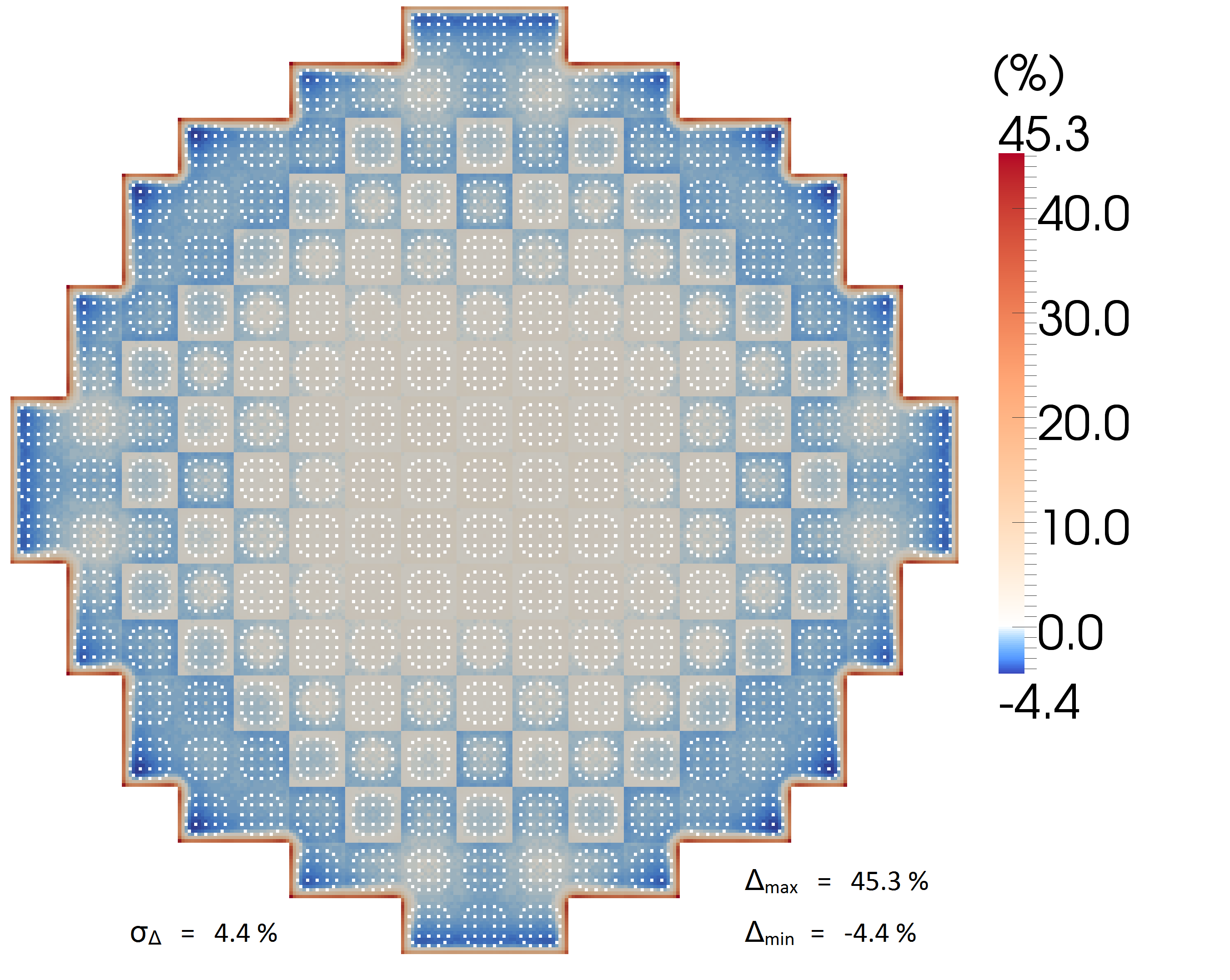}
  \caption{Power discrepancy distribution pin by pin after optimization}
  \label{fig:ComparaisonCC_4grSPN_ADAO}
 \end{center}
\end{figure}

The optimization performed by our computational scheme has a significant effect on the power discrepancies distribution. In fact, \mbox{$\text{E}_{\Delta_{\text{abs}}}$=10.12\%} and \mbox{$\sigma_{\Delta}$=11.92\%} initially and respectively \mbox{0.66\%} and \mbox{0.82\%} after optimization. Moreover, the assemblies at the interface core/reflector and the assembly 6 are better described than with the diffusion solver. The modeling choices in this section enable to better deal with the heterogeneous regions. The results are similar to those obtained with the optimized Lefebvre-Lebigot reflector presented in Section 4.1. Given that the core is modeled pin by pin, the power discrepancy distribution pin by pin is also presented, in Figure~\ref{fig:ComparaisonCC_4grSPN_ADAO}.

This figure still shows that the maximum discrepancies are located in the last rows of pins at the interface core/reflector and are very high (\mbox{$\Delta_{\text{max}}$ = 45.3 \%
}). However, these discrepancies are less important than for diffusion calculation (\mbox{$\Delta_{\text{max}}$ = 78.9 \%} on Figure~\ref{fig:RFPApresOpt-4gr}). This is probably caused by the homogeneous by assembly modeling of the reflector. The assemblies at the interface core/reflector are composed of various materials in reality and the homogeneous modeling seems to cause this wrong description of the reflector at the interface. The discrepancy distribution is very homogeneous in the center of the core (the grey zone on Figure~\ref{fig:ComparaisonCC_4grSPN_ADAO}), and the discrepancies shown are very low.

\subsection{8-group calculations}
\begin{figure}[H]
 \begin{center}
  \includegraphics[scale=0.28]{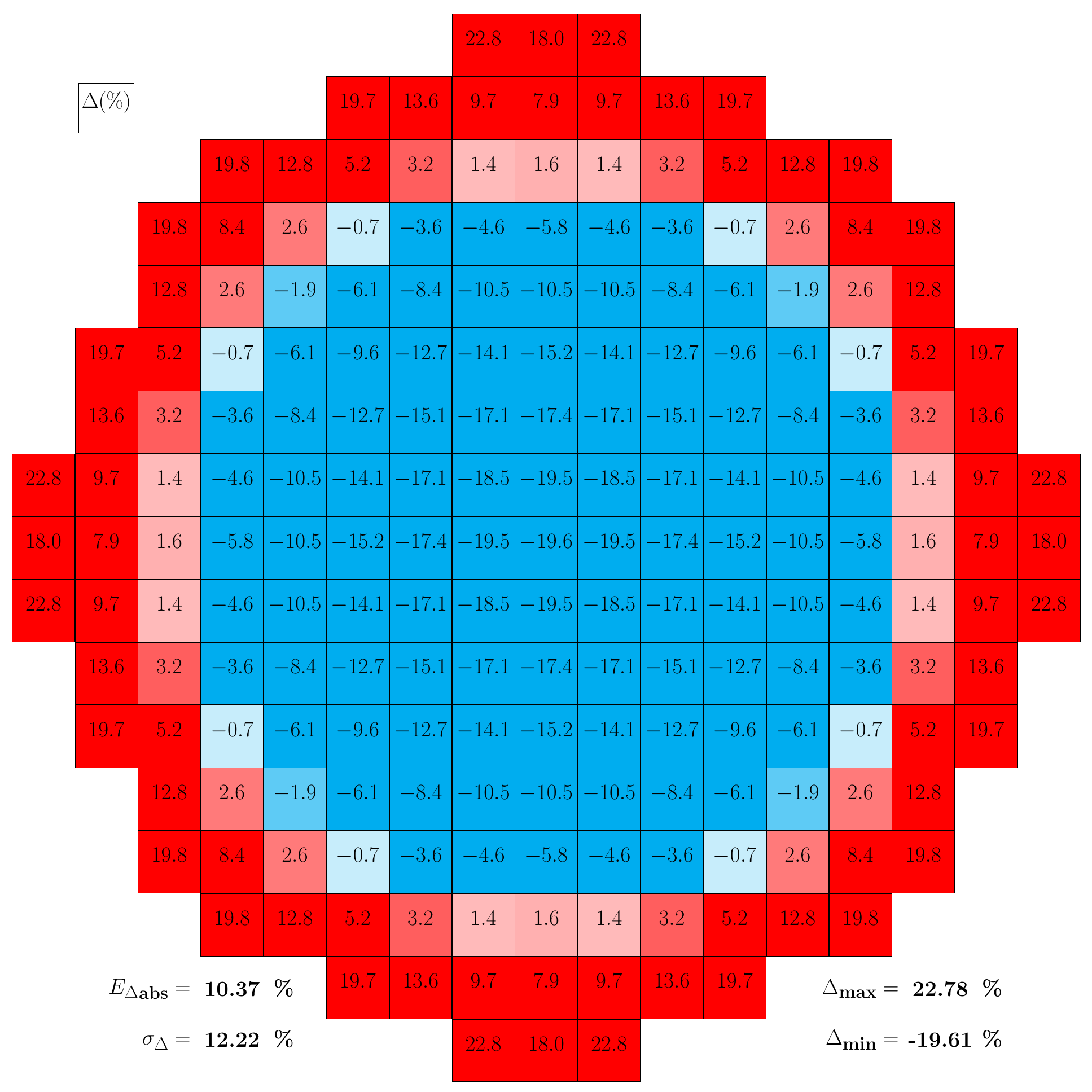}
  \caption{Power discrepancy distribution (before opt.)}
  \label{fig:puissanceCOCDRAG_8grSPN_D11D16_AvantOpt}
 \end{center}
\end{figure}

\begin{figure}[H]
 \begin{center}
  \includegraphics[scale=0.28]{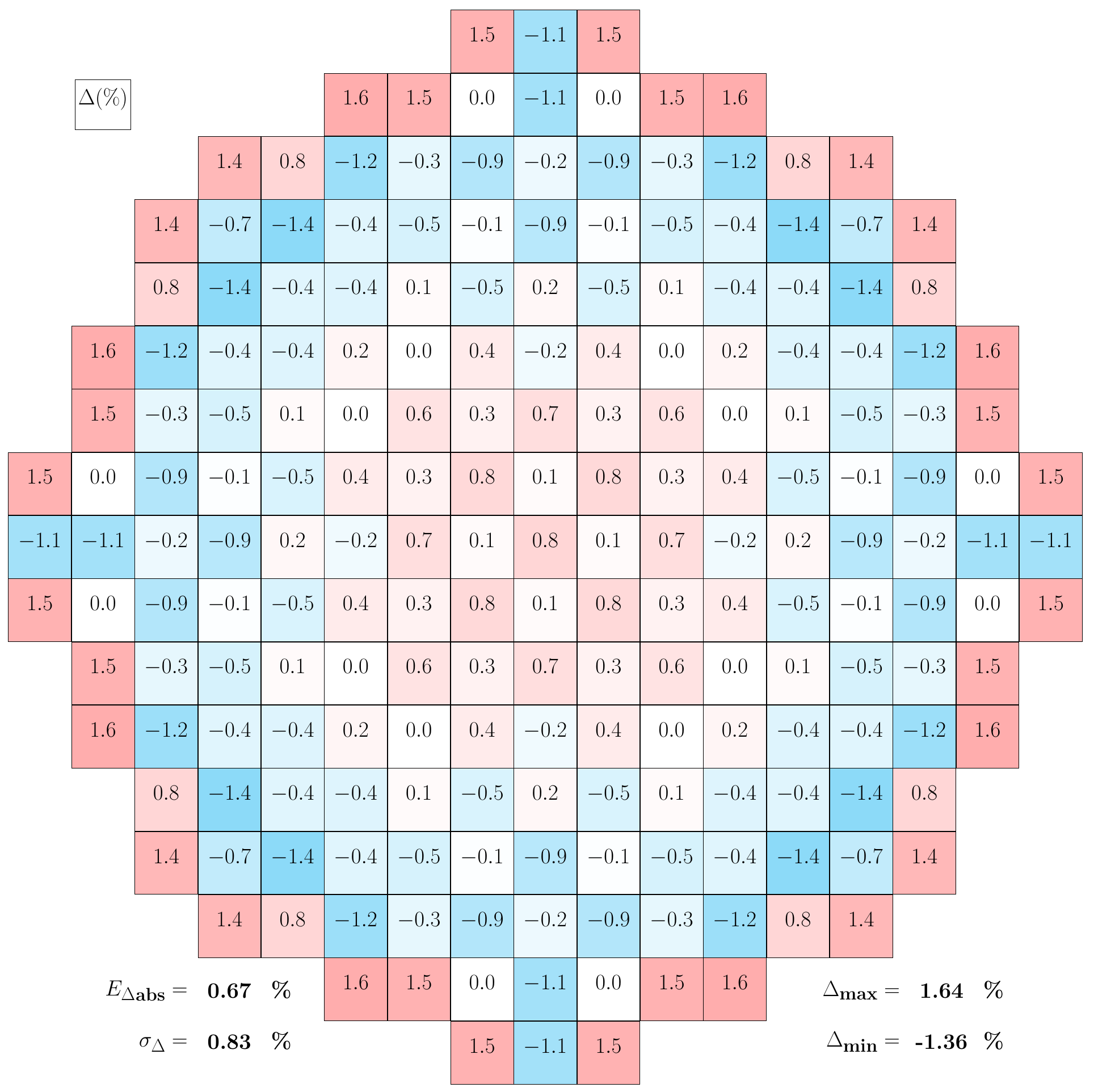}
  \caption{Power discrepancy distribution (after opt.)}
  \label{fig:puissanceCOCDRAG_8grSPN_D11D16_ADAO}
 \end{center}
\end{figure}

In this section, 8-group calculations are performed. The modeling choices are still the same than in Section 5.1. The discrepancies distributions before and after optimization are presented in Figure~\ref{fig:puissanceCOCDRAG_8grSPN_D11D16_AvantOpt} and Figure~\ref{fig:puissanceCOCDRAG_8grSPN_D11D16_ADAO}. Figure~\ref{fig:puissanceCOCDRAG_8grSPN_D11D16_ADAO} shows that refining the energy mesh from 4 to 8 groups doesn't have a significant impact on the assembly discrepancy distribution (\mbox{$\text{E}_{\Delta_{\text{abs}}}$=0.66\%} and \mbox{$\sigma_{\Delta}$=0.82\%} for 4-group calculations and respectively \mbox{0.67\%} and \mbox{0.83\%} for 8-group calculations). However, slight improvement are observed. The assembly 6 shows a lower discrepancy for 8-group calculation (\mbox{-0.9 \%} against \mbox{-1.2 \%} for 4-group calculation). Moreover, the difference of the discrepancies in two consecutive assemblies are lower (\mbox{$\Delta$ = 0.8 \%} in assembly 1 and \mbox{0.1 \%} in assembly 2 for 8-group calculations, and \mbox{$\Delta$ = 0.8 \%} in assembly 1 and \mbox{-0.1 \%} in assembly 2 for 4-group calculations).

\begin{figure}[H]
 \begin{center}
  \includegraphics[scale=0.1]{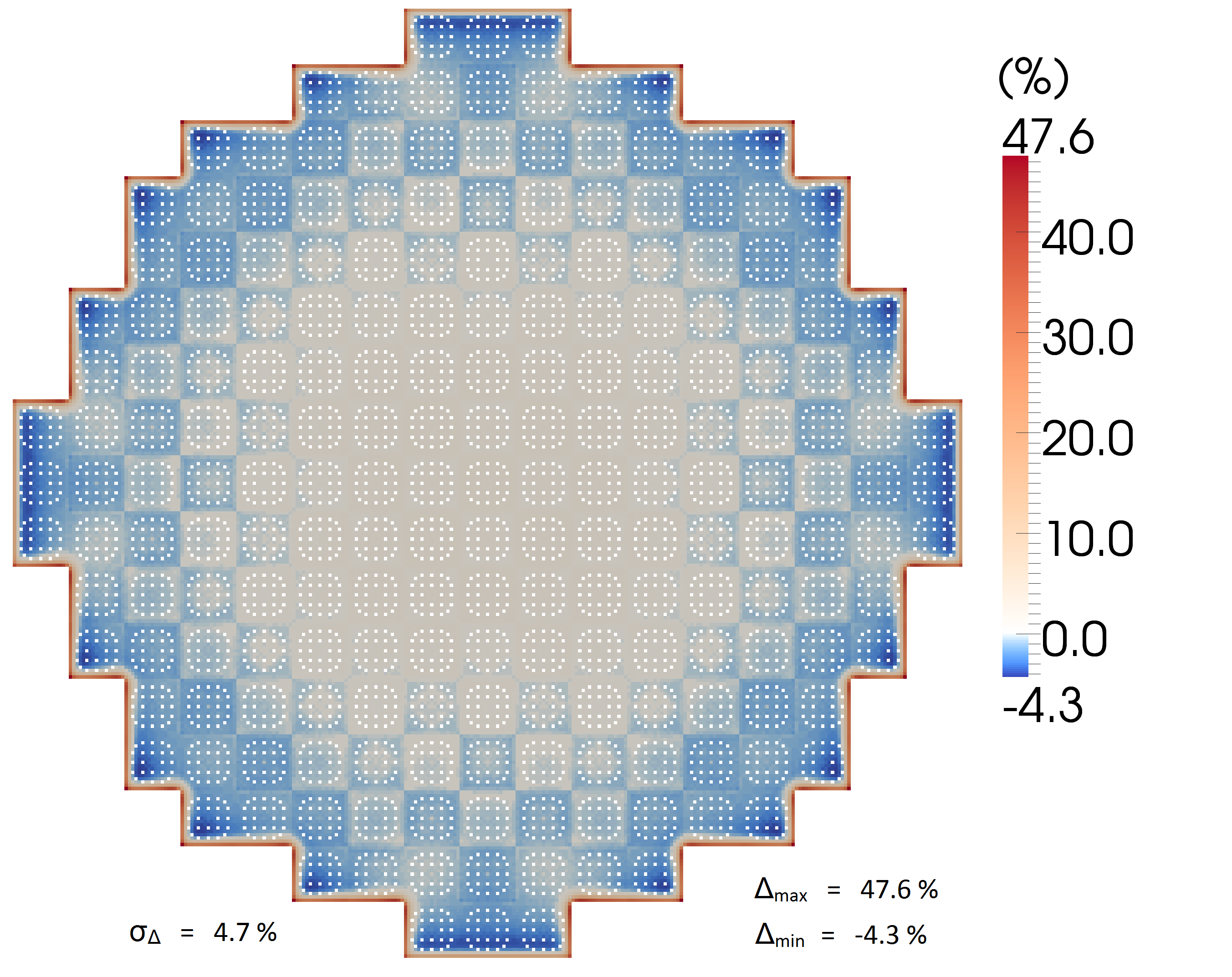}
  \caption{Power discrepancy distribution pin by pin after optimization}
  \label{fig:ComparaisonCC_8grSPN_ADAO}
 \end{center}
\end{figure}

Here again, the pin-by-pin power discrepancies distribution is presented in Figure~\ref{fig:ComparaisonCC_8grSPN_ADAO}.

This figure shows that the maximum discrepancies are still located in the last rows of pins at the interface core/reflector and are still very high (\mbox{$\Delta_{\text{max}}$ = 47.6 \%}). The pin by pin visualization shows additional information: the homogeneous zone at the center of the core is more important for 8-group calculations than for 4-group calculation (the grey zone on Figure~\ref{fig:ComparaisonCC_8grSPN_ADAO}). 

\subsection{Conclusion on $\text{SP}_{\text{3}}$ calculations}
In this section, we have presented the results of our computational scheme fro 4-group and 8-group $\text{SP}_{\text{3}}$ calculations. The results show that the $\text{SP}_{\text{3}}$ operator has a strong impact on the discrepancies distribution compared to the diffusion operator. The power discrepancies distributions are clearly improved compared to Section 4.2, and the results are similar to those obtained with the Lefebvre-Lebigot reflector. Though, very high discrepancies are observed in the last rows of pins at the interface core/reflector, due to the homogenization performed that doesn't represent the reflector's heterogeneities.

\section{Discussion and Conclusion}

In this paper, we have designed and fully validated a computational scheme able to compute equivalent multi-group 2D reflectors, for diffusion or $\text{SP}_{\text{N}}$ operators. The validation has been carried out using the OPTEX reflector model developed at EPM, and enabled to identify modeling choices. In all this study, all reflectors have been designed by optimizing the fast reflector parameter (diffusion coefficient or P-1 weighted macroscopic cross-section) in each region of the reflector. Moreover the reflector has been initialized with the Lefebvre-Lebigot method when possible, and with collapsing/homogenization of an APOLLO2 reference calculation in other cases. As a result, we have designed reflector for diffusion calculation with 2 and 4 energy groups, and for $\text{SP}_{\text{3}}$ calculation, with 4 and 8 energy groups. All the results show that our computational scheme causes an improvement of the power discrepancies distribution with the reference distribution. This computational scheme is then an interesting tool to design 2-group and multi-group 2D reflectors in PWRs.
Though, it is important to note that when using a initial reflector computed by direct collapsing/homogenization of the reference, the discrepancies at the interface core/reflector are still very important. These high discrepancies are probably caused by the the following reasons:
\begin{enumerate}
  \item[$\bullet$] The reflector is represented as homogeneous blocks, each one of the size of a fuel assembly. This modeling doesn't reflect the reflector's heterogeneities. It would be interesting to design an hybrid modeling of the reflector, pin-by-pin near the core/reflector interface, and with homogeneous assemblies far from the interface. An other solution would be to create a meshing of the reflector with macro groups of pins that represent the several materials that compose it.\
  \item[$\bullet$] The infinite medium calculation of the core seem to reach its limit at the interface core/reflector. In the center of the core, where the enrichment of the assemblies are very close, this approximation makes sense, but for assemblies close to the interface, it would be interesting to take into account the environment of these assemblies in the cross-section calculation.\
  \item[$\bullet$] At the end of the reference MOC calculation performed with APOLLO2, the 26-group diffusion coefficients of each reflector block are computed using the simple formula $\text{D} = \frac{1}{3\Sigma_{\rm \text{tr}}}$. This is a crude approximation that could certainly be improved in a future study.\
\end{enumerate}
These limitations are to be explored in future work.

\clearpage 

\renewcommand\bibname{References}
\bibliography{Article_Annals}
\bibliographystyle{elsarticle-harv}

\end{document}